\documentclass[aps,preprint,nofootinbib,floatfix]{revtex4-1}
\usepackage{graphicx,color}
\usepackage[latin1]{inputenc}
\usepackage{amsmath,amssymb}
\usepackage{hyperref}
\usepackage{epstopdf}
\usepackage{geometry}                
\usepackage[normalem]{ulem}
\def\be{\begin{equation}}
\def\ee{\end{equation}}
\def\bea{\begin{eqnarray}}
\def\eea{\end{eqnarray}}
\newcommand{\gev}{~{\rm GeV}}

\begin{document}
\preprint{OSU-HEP-16-03}
\title{The search for electroweak-scale right-handed neutrinos and mirror charged leptons through like-sign dilepton signals}
\author{Shreyashi Chakdar$^{2}$}
\email{chakdar@virginia.edu}
\author{K. Ghosh$^{4}$}
\email{kirti.gh@gmail.com}
\author{V. Hoang$^2$}
\email{vvh9ux@virginia.edu}
\author{P. Q. Hung$^{2,3}$}
\email{pqh@virginia.edu}
\author{S. Nandi$^{1}$}
\email{s.nandi@okstate.edu}

\affiliation{$^1$Department of Physics and Oklahoma Center for High Energy Physics,
Oklahoma State University, Stillwater, OK 74078-3072, USA \\
$^2$Department of Physics, University of Virginia, Charlottesville, VA 22904-4714, USA\\
$^3$Center for Theoretical and Computational Physics, Hue University College of Education, Hue, Vietnam.\\
$^4$Department of Physics and Astrophysics, University of Delhi, Delhi 110007, India.
}


\begin{abstract}
The existence of tiny neutrino masses at a scale more than a million times smaller than the lightest charged fermion mass, namely the electron, and their mixings  can not be explained within the framework of the exceptionally successful Standard Model(SM). Several mechanisms were proposed to explain the tiny neutrino masses, most prominent among which is the so-called seesaw mechanism. Many models were built around this concept, one of which is the EW-scale $\nu_R$ model.  In this model, right-handed neutrinos are {\em fertile} and their masses are connected to the electroweak scale $\Lambda_{EW}\sim246 \gev$. It is these two features that make the search for right-handed neutrinos at colliders such as the LHC feasible. The EW-scale $\nu_R$ model has new quarks and leptons of opposite chirality at the electroweak scale (for the same SM gauge symmetry $SU(2)_W \times U(1)_Y$) compared to what we have for the Standard Model. 
With suitable modification of the Higgs sector, the EW-scale $\nu_R$ model satisfies the electroweak precision test and,
also the constraints coming from the observed 125-GeV Higgs scalar.
Since in this model, the mirror fermions are required to be in the EW scale, these can be produced at the LHC giving final states with a very low background from the SM. One such final state is the same sign dileptons with large missing $p_T$ for the events. In this work, we explore the constraint provided by the $8$ TeV data, and prospect of observing this signal in the $13$ TeV runs at the LHC. Additional signals will be the presence of displaced vertices depending on the smallness of the Yukawa couplings of the mirror leptons with the ordinary leptons and the singlet Higgs  present in the model. Of particular importance to the EW-scale $\nu_R$ model is the production of $\nu_R$ which will be a direct test of the seesaw mechanism at collider energies.

\end{abstract}


\maketitle

\section{Introduction}

The Standard Model (SM) based on the gauge symmetry $SU(3)_C \times SU(2)_L \times U(1)$  has been remarkably successful to explain all the phenomena from very low energies to all the way at the highest energy Large Hadron Collider (LHC). 
However, there are two experimental observations (among many such as the mass hierarchies among the quarks and the leptons, the CKM matrix, the PMNS matrix,...)
which the SM can not explain. These are the existence of the dark matter in the universe, and the tiny non-zero masses of the neutrinos and their mixings. The SM has no candidate for the dark matter. Tiny neutrino masses can be generated using so called Weinberg operator which is  
of dimension five and suppressed by a scale M \cite{Weinberg}. However, if SM is the ultimate theory, then scale M is the Planck's scale. In this case, the neutrino masses generated are three or four orders of magnitude smaller than the observed ones.  Several ideas have been proposed to generate tiny neutrino masses involving physics beyond the SM. The most elegant one is the so called see-saw \cite{See-Saw} mechanism. 
Several models have been constructed around the seesaw mechanism. A popular version is one in which the SM gauge symmetry is extended to higher gauge symmetry, such as left-right extension \cite{L-R} of the SM, or $SO(10)$ grand unified theory (GUT). Other popular versions come under the names: Type-I seesaw where extra heavy fermion singlets, the right-handed neutrinos, are added to the SM \cite{See-Saw}, Type-II seesaw where extra scalar triplets are added \cite{type2} and Type-III seesaw where extra fermion triplets are added \cite{type3}. Also there are models based on the so-called "low scale seesaw" where the right-handed neutrinos have masses below the electroweak scale \cite{lowscale}. There exists models for tiny neutrino masses which are not based on the seesaw mechanism such as radiative neutrino mass generation \cite{Zee-Babu} by extending the SM particle content to include extra Higgs multiplets, and choosing the masses of these additional Higgs bosons and the associated Yukawa couplings suitably, and an extension the SM to include three RH neutrinos and a 2nd Higgs doublet having VEV at the KeV scale \cite{G-N}. Lepton number conservation is imposed. An exhaustive list of neutrino mass models is beyond the scope of this manuscript.

A seesaw model, called the "Electroweak $\nu_R$ model" (or EW-sale $\nu_R$ model for short), was proposed \cite{Hung} in which the gauge symmetry  $SU(2)_W \times U(1)$ is the same as that of the SM, supplemented by an $A_4$ discrete symmetry. The distinguishing feature of this model is the fact that right-handed neutrinos are {\em fertile} (as opposed to {\em sterile}), unlike Type-I seesaw models.
 In fact, the model  has  several additional fermion multiplets  (for every LH doublets, there are RH doublets, and for every RH singlets, there are left handed singlets (called the mirror) fermions at the electroweak scale). The model also has one additional Higgs doublet, called the mirror doublet, two triplets, and four singlets. In this model, the seesaw mechanism gives rise to tiny neutrino masses with the RH neutrino in the EW scale, and the neutrinos are Majorana type. The right-handed neutrinos interact with the electroweak W's and Z bosons (and hence the use of the adjective "fertile"), a feature which allows for their search at colliders and directly test the seesaw mechanism.The rationale for this particle content is explained in the Review section.

The idea of mirror quarks and mirror leptons is not new. This was considered by Lee and Yang back in $1950$'s \cite{LY}. Their argument was that the LH quarks and leptons having weak interaction, while the the RH handed ones do not, is not symmetric in nature. So they speculated that there may exist fermions of exactly opposite chirality, and those have not been observed experimentally because those are very heavy. However, now we know that  adding these mirror fermions with the gauge symmetry of the SM does not satisfy the precision EW test involving the S parameter. These will contribute positively too much to the S parameter. This is circumvented by adding Higgs triplets to the models which gives large negative contribution and thus satisfying the S parameter constraints \cite{Hung}. In fact, the complex Higgs triplet was introduced in the EW-scale $\nu_R$ model for another purpose: It gives a Majorana mass to {\em fertile} right-handed neutrinos which is proportional to the electroweak scale $\Lambda_{EW} \sim 246 \gev$. The fact that its contribution to the S-parameter can largely offset that of the right-handed mirror fermions in a large region of parameter space is an unexpected bonus of the EW-scale $\nu_R$ model \cite{Hung}, \cite{hung2}.  One of the major feature of the model is that since the symmetry group is just the SM, both Higgs doublets have VEVs in the EW scale. Thus to satisfy perturbative constraint, all the particles, all the mirror fermion, as well as the $\nu_R$ have masses in the EW scale, and less than a TeV. Thus these particles can be produced at the LHC, and  the ensuing final sates can be looked for in search of new physics signals. 

Lastly, the electroweak phase transition is non-perturbative in nature. One approach for studying non-perturbative phenomena is that of lattice gauge theory. It is well-known that one cannot put a chiral gauge theory such as the SM on the lattice because of the loss of gauge invariance. Ref.~\cite{montvay} proposed the introduction of mirror fermions in order to achieve a gauge-invariant formulation of the SM on the lattice. The mirror fermions of the EW-scale $\nu_R$ model fits that bill.

In a previous work \cite{CGHHN}, we discussed the signals of the electroweak $\nu_R$ model at the LHC arising from the pair productions of the mirror quarks as developed in \cite{Hung} and the subsequent extensions of the model satisfying all the constraints. We found that for the prompt decays of the mirror quarks plus almost massless neutral scalar (present in the model), the mass of the lightest mirror quark as low as $600$ GeV is allowed from the $8$ TeV data assuming  this branching ratio to be $100\%$. And if this decay branching  is $50\%$ or less, their is no bound from the LHC $8$ TeV data. We also calculated the final state signals for the $13$  TeV LHC, and found that the reach for the lightest mirror quark can be as large as $\simeq 700$ GeV with $ \simeq 100 fb^{-1}$ luminosity. 

In this work, we explore the new physics that might arise from the pair productions of the mirror leptons at the LHC. The mirror lepton production cross sections are much lower compared to the mirror quarks, because those are produced via EW interaction, whereas mirror quarks are produced via the strong interaction. However, the pair production of mirror leptons, such as $\nu_R^M  \nu_R^M$ and $\nu_R^M e_R^M$ give rise to high $p_T$ same sign dileptons and trileptons ($++-$ or $- - +$) with large missing energy. The SM background for the such final state are very small, and signals, if found, will stand well above the background. 
We make prediction for the signal in the $13$ TeV data as a function of the mirror neutrino mass for several values of the charged lepton masses, as well as the background (which is small). Since the mirror lepton masses have to smaller than a TeV, there is a good chance that new physics may be discovered in the upcoming runs at the LHC if this model is realized in nature.

Our presentation below is organized as follows.  In section II, we review the model and the formalism in some detail. This include the gauge sector, fermion sector and the scalar sector, and the neutrino masses. In section III, we discuss the precision EW constraint for the model. In section IV, we discuss the constraints coming from the available $125$ GeV Higgs data. In section V, we discuss the Yukawa interaction in the model. 
The collider signatures of pair productions of mirror leptons in the framework of $EW-\nu_R$ model was discussed Section VI. Section VII contains our conclusions and discussions.


\section{ The model, formalism and the existing constraints}

In this section, we will summarize the essential features of the EW-scale $\nu_R$ model \cite{Hung} with a particular emphasis on the meaning and use of the particle content of the model and its comparison with the well-known Left-Right symmetric model \cite{LR} in its various versions. 

As stipulated in the introduction, the rationale for the construction of the EW-scale $\nu_R$ model was to "bring down" the energy scale of the seesaw mechanism to the electroweak scale by making right-handed neutrinos "non-sterile" or "fertile". This has the clear advantage of being able to test the concept of the seesaw mechanism by directly searching for those fertile right-handed neutrinos at colliders such as the LHC and/or at future colliders such as the ILC. As Ref.~\cite{Hung} has shown, in order to realize this scenario, it is necessary to introduce new degrees of freedom beyond those of the SM: right-handed $SU(2)$ mirror quark and lepton doublets, left-handed mirror quark and lepton $SU(2)$ singlets, two Higgs triplets (one real and one complex), two Higgs doublets, and four Higgs singlets. The gauge group is identical to that of the SM. Before writing down explicitly the particle content of the model, we would have to address the usual concern that one may have whenever one goes beyond the Standard Model: Are there too many extra degrees of freedom and what do they accomplish? (It goes without saying that this kind of concern applies to {\em all} BSM models.) It is for this purpose that this section is devoted to the description of the EW-scale $\nu_R$ model and its comparison with the popular Left-Right Symmetric model $SU(2)_L \times SU(2)_R \times U(1)_{B-L}$. This comparison with the very popular LR model is simply for the purpose of showing that the scalar sector of the EW-scale $\nu_R$ model is not overly complicated. We first list the particle content and subsequently discuss what these particles are used for.

I) {\bf The gauge sector}: 

\begin{itemize}

\item {\bf Gauge group of the EW-scale $\nu_R$ model}:

\be
SU(3)_C \times SU(2)_W \times U(1)_Y
\ee

\item {\bf Gauge group of Left-Right models}:

\be
SU(3)_C \times SU(2)_L \times SU(e)_R \times U(1)_{B-L}
\ee

\item 
As it will be reviewed below, the EW-scale $\nu_R$ model has several scalar multiplets (two doublets and two triplets) whose VEV's contribute to the electroweak scale $\Lambda_{EW} \sim 246\gev$, namely $v_{2}^2 +v_{2M}^2 + 8 v_{M}^2= v^2 \approx (246 \gev)^2$. The {\em effective breaking scale} of the EW$\nu_R$ model is $\Lambda_{EW} \sim 246\gev$ which is the maximum scale of the model. The W and Z masses are directly proportional to $\Lambda_{EW}$. Similarly, the L-R model has several VEV's coming from scalar doublets and triplets for both $SU(2)_L$ and $SU(2)_R$. This results in two effective breaking scales usually characterized by the masses of $W_L$ and $W_R$ with the latter currently bounded from below by 3 TeV. 
Finally, as we shall see below, the various scalar multiplets, beside their contributions to the electroweak scale, play an important role in fermion masses. 

\end{itemize}

II) {\bf The Fermion sector}:

\begin{itemize}


\item {\bf Fermion $SU(2)_W$ doublets} ($M$ refers to mirror fermions):

SM: $l_L = \left(
	  \begin{array}{c}
	   \nu_L \\
	   e_L \\
	  \end{array}
	 \right)$; Mirror: $l_R^M = \left(
	  \begin{array}{c}
	   \nu_R^M \\
	   e_R^M \\
	  \end{array}
	 \right)$. \
	 
	 Notice that right-handed neutrinos are "fertile" in the EW-scale $\nu_R$ model because they are now parts of right-handed lepton doublets. How heavy they can be will be the subject of the section on Majorana masses below. 
	 
SM: $q_L = \left(
	  	 \begin{array}{c}
	   	  u_L \\
	     	  d_L \\
	  	\end{array}
	 	\right)$; Mirror: $q_R^M = \left(
	  	 \begin{array}{c}
	   	  u_R^M \\
	     	  d_R^M \\
	  	\end{array}
	 	\right)$.

\item {\bf Fermion $SU(2)_W$ singlets}:

SM: $e_R; \ u_R, \ d_R$; Mirror: $e_L^M; \ u_L^M, \ d_L^M$

\item {\bf Fermions in Left-Right models}:

$SU(2)_L$:  $l_L = \left(
	  \begin{array}{c}
	   \nu_L \\
	   e_L \\
	  \end{array}
	 \right)$; 
	 
$SU(2)_R$: $l_R = \left(
	  \begin{array}{c}
	   \nu_R \\
	   e_R\\
	  \end{array}
	 \right)$. \
	 
$SU(2)_L$:  $q_L = \left(
	  	 \begin{array}{c}
	   	  u_L \\
	     	  d_L \\
	  	\end{array}
	 	\right)$; 
		
$SU(2)_R$: $q_R= \left(
	  	 \begin{array}{c}
	   	  u_R \\
	     	  d_R\\
	  	\end{array}
	 	\right)$.
		
\end{itemize}		

III) {\bf The scalar sector}:

\begin{itemize}

\item{The scalar sector in the EW-scale $\nu_R$ model}
 
	 a) Higgs doublets:
	 
	  $\Phi_2=\left(
	 			\begin{array}{c}
				\phi_{2}^+ \\
				\phi_{2} ^0 \\
				\end{array}
				\right)$ 		
with $\langle \phi_{2}^0 \rangle = v_2/\sqrt{2}$. 

In the original version \cite{Hung}, this Higgs doublet couples to both SM and mirror fermions. An extended version was proposed \cite{hung3} in order to accommodate the 125-GeV SM-like scalar and, in this version, $\Phi_2$ only couples to SM fermions while another doublet $\Phi_{2M}$ whose VEV is $\langle \phi_{2M}^0 \rangle = v_{2M}/\sqrt{2}$ couples only to mirror fermions. 

$\Phi_{2M}=\left(
	 			\begin{array}{c}
				\phi_{2M}^+ \\
				\phi_{2M} ^0 \\
				\end{array}
				\right)$ 		
with $\langle \phi_{2M}^0 \rangle = v_{2M}/\sqrt{2}$. 

	 b) Higgs triplets:
	i) Complex triplet: $\widetilde{\chi} \ (Y/2 = 1)  = \frac{1}{\sqrt{2}} \ \vec{\tau} . \vec{\chi} = 
	  \left(
	  \begin{array}{cc}
	    \frac{1}{\sqrt{2}} \chi^+ & \chi^{++} \\
	    \chi^0 & - \frac{1}{\sqrt{2}} \chi^+\\
	   \end{array}
		  \right)$ with $\langle \chi^0 \rangle = v_M$.
			
	ii) Real triplet: $\xi \ (Y/2 = 0)$ in order to restore Custodial Symmetry with $\langle \xi^0 \rangle = v_M$.
			
	The VEVs are given by:
	
	$v_{2}^2 +v_{2M}^2 + 8 v_{M}^2= v^2 \approx (246 \gev)^2$

	c) Four Higgs singlets: This was needed to construct neutrino mass matrices within the framework of an $A_4$ non-abelian discrete symmetry \cite{hung4}. 	
		
\item{The minimal scalar sector in the Left-Right model} 

a) Two complex Higgs triplets: $\Delta_R=(1,3)$ and $\Delta_L=(3,1)$ under $SU(2)_L \times SU(2)_R$. It is generally assumed that $\langle \Delta_L \rangle =v_L \ll \Lambda_{EW}$ in order to satisfy $\rho \approx 1$. Furthermore, $\langle \Delta_R \rangle =v_R > 3 TeV $.

b) A bi-doublet $\Phi = (2,2)$ which is equivalent  to {\em two} SM doublets.

\end{itemize}

IV) {\bf  The role of the scalar sector }

Here we summarize the salient parts of the Yukawa interactions in the EW-scale $\nu_R$ model followed by the crucial roles of the gauge Higgs triplets in ensuring the agreement between the EW-scale $\nu_R$ model and electroweak precision data. 

\begin{itemize}
		
\item {\bf Dirac and Majorana Neutrino Masses, charged fermion masses in the EW-scale $\nu_R$ model}
For simplicity, from hereon, we will write $\nu_R^M$ simply as $\nu_R$.

a) Majorana Neutrino Masses:
 
 The main point of \cite{Hung} is the fact the right-handed neutrinos are now non-sterile and are expected to acquire a mass proportional to the electroweak breaking scale. This is achieved by

\bea
\label{majorana}
L_M &= &g_M \, l^{M,T}_R \ \sigma_2 \ \tau_2 \ \tilde{\chi} \ l^M_R \\ \nonumber
&= &g_M \ \nu_R^T \ \sigma_2 \ \nu_R \ \chi^0 - \dfrac{1}{\sqrt{2}} \ \nu_R^T \ \sigma_2 \ e_R^M \ \chi^+ \\ \nonumber
&&- \dfrac{1}{\sqrt{2}} \ e_R^{M,T} \ \sigma_2 \ \nu_R \ \chi^+ + e_R^{M,T} \ \sigma_2 \ e_R^M \ \chi^{++} \,.
\eea
From (\ref{majorana}), we obtain the Majorana mass $ M_R = g_M v_M $. Since $\nu_R$'s interact with the Z-boson, The Z-width requires that $ M_R = g_M v_M > M_Z/2$ implying that $v_M > 46 \gev$. Such a "large" VEV would destroy the tree-level relationship $\rho= M_{W}^2/ M_{Z}^2 \cos ^2 \theta_W =1$ if not for the presence of the real triplet $\xi \ (Y/2 = 0)$ as we will mention below \cite{Hung}.

Since the EW-scale $\nu_R$ model contains {\em non-sterile} right-handed neutrinos with masses $\propto O(\Lambda_{EW} \sim 246 \gev)$, one expects to be able to produce them at the LHC through elementary processes such as $\bar{q} q \rightarrow Z \rightarrow \nu_R \nu_R$ and $\bar{u} d \rightarrow W^{-} \rightarrow \nu_R e_{R}^{M,-}$, where $e_{R}^{M,-}$ denotes a generic mirror charged lepton. The production cross sections are typically electroweak cross sections and distinct signatures are like-sign dileptons \cite{Hung}. Detailed calculations of such processes will be the subject of our manuscript.

It is important to note here that in the original version \cite{Hung}, a global symmetry denoted by $U(1)_M$ was assumed under which the mirror right-handed doublets and left-handed singlets transform as $(l_R^{M}, e_L^{M}) \rightarrow e^{\imath \theta_M} (l_R^{M}, e_L^{M})$ and the triplet and singlet Higgs fields transform as $\tilde{\chi} \rightarrow e^{-2\imath \theta_M} \tilde{\chi}$, $\phi_S \rightarrow e^{-\imath \theta_M} \phi_S$, with all other fields being singlets under $U(1)_M$. With this transformation, a coupling similar to Eq  (\ref{majorana}) is forbidden for the SM leptons and hence there is no Majorana mass for left-handed neutrinos at tree level. It was also shown in \cite{Hung} that the Majorana mass for left-handed neutrinos can arise at one loop but is much smaller than the light neutrino mass and thus can be ignored.

b) Dirac Neutrino Mass:

As described in \cite{Hung}, Dirac neutrino masses are obtained by the mixed coupling between SM and Mirror leptons with Higgs singlets. A generalization of a single Higgs singlet to four \cite{pqtrinh} (3+1 representations of $A_4$) was proposed to accommodate neutrino masses and mixings. For this purpose of review, we only need to show a generic coupling.

The singlet scalar field $\phi_S$ couples to fermion bilinears as
\bea
\label{dirac}
L_S &=& g_{Sl} \,\bar{l}_L \ \phi_S \ l_R^M + h.c.\\  \nonumber
         &= &g_{Sl} (\bar{\nu}_L \ \nu_R \ + \bar{e}_L \ e_R^M) \ \phi_S + h.c. \,.
\eea
From (\ref{dirac}),  we get the Dirac neutrino masses $m_\nu^D = g_{Sl} \ v_S $. The seesaw mechanism and phenomenological constraints give $m_\nu l \sim (m_\nu^D)^2/M_R < O(eV)$, implying $g_{Sl} \ v_S  < O(100 keV)$. The physics involving the singlet scalars were discussed in \cite{pqtrinh} and \cite{hung4}.

c) Masses of charged leptons and quarks:

As shown in \cite{Hung} and in the extension of the EW-scale $\nu_R$ model \cite{hung3}, SM quarks and charged leptons obtain masses by coupling to the doublet $\Phi_2$ and mirror quarks and charged leptons obtain theirs by coupling to $\Phi_{2M}$. 

\item{Dirac and Majorana neutrino masses, charged fermion masses in L-R models}

a) Majorana masses:

Right-handed neutrinos which belong to doublets of $SU(2)_R$ along with the SM right-handed charged leptons obtain Majorana masses from the VEV of $\Delta_R$, namely $\langle \Delta_R \rangle =v_R > 3 TeV $. Typically, $M_R \sim O(v_R)$. As a result, these Majorana masses are large. In addition, right-handed neutrinos in L-R models can only be produced through the exchanges of $W_R$ and $Z_R$ whose masses are constrained to be above 3 TeV. Because of the high mass constraints on $W_R$ and $Z_R$, one expects much smaller production cross sections than those of the EW-scale $\nu_R$ model. 

b) Dirac masses:

Dirac neutrino masses in the L-R models are obtained by coupling to the Higgs bi-doublet so that $m_D \sim O(v_L \sim \Lambda_{EW})$. Similarly, charged lepton and quark masses are obtained also by coupling to the {\em same} Higgs bi-doublet. 

Let us recall from the previous section that Dirac neutrino masses in the EW-scale $\nu_R$ model come from the coupling to the Higgs {\em singlets} while those of the SM charged lepton and quarks and their mirror counterparts are gotten from the coupling to $\Phi_{2}$ and $\Phi_{2M}$ respectively. It is this difference in coupling that \cite{pqtrinh} exploited in showing the difference between the PMNS and CKM matrices.

\item{The scalar contributions to electroweak radiative corrections}

Since the mini-review of electroweak precision constraints on the EW-scale $\nu_R$ model will be given below, we just mention in this section the salient points of the triplet scalar contributions to the precision parameters such as S, T, U. It was noticed in \cite{Hung} and \cite{Adibzadeh:2007hz} that triplet Higgs representations can give {\em large negative contributions} to the S-parameter. In fact, any model containing Higgs triplets can give such a negative contribution to the S-parameter in a large region of parameter space. As can be seen in \cite{Adibzadeh:2007hz} and subsequently in \cite{hung2}, {\em fine tuning} is required if one desires to have a very small contribution to the S-parameter coming solely from the Higgs triplet. Such a fine-tuning disappears if the {\em negative} contribution from the Higgs triplet cancels against a {\em positive} contribution from an extra fermion sector. This is the case with the EW-scale $\nu_R$ model \cite{hung2} where the positive contribution coming from mirror fermion doublets cancels against the negative contributions coming from the scalar sector, in particular the Higgs triplet.  This is summarized below.                                  

\end{itemize}
 

The next two sections are reviews of the electroweak precision constraints on the EW $\nu_R$ model \cite{hung2} as well as the constraint coming from the 125-GeV scalar. These sections and the one above are necessary to introduce the model to readers who are not familiar with it and we include similar reviews in all related papers.

\section  {Electroweak precision constraints on the EW $\nu_R$ model \cite{hung2}}

The presence of mirror quark and lepton $SU(2)$-doublets can, by themselves, seriously affect the constraints coming from electroweak precision data. As noticed in \cite{Hung}, the positive contribution to the S-parameter coming from the extra right-handed mirror quark and lepton doublets could be partially cancelled by the negative contribution coming from the triplet Higgs fields. Ref.~\cite{hung2} has carried out a detailed analysis of the electroweak precision parameters S and T and found that there is a large parameter space in the model which satisfies the present constraints and that there is {\em no fine tuning} due to the large size of the allowed parameter space. It is beyond the scope of the paper to show more details here but a representative plot would be helpful. Fig. 1 shows the contribution of the scalar sector versus that of the mirror fermions to the S-parameter within 1$\sigma$ and 2$\sigma$.
\begin{figure}[t]
\centering
   \includegraphics[scale=0.35]{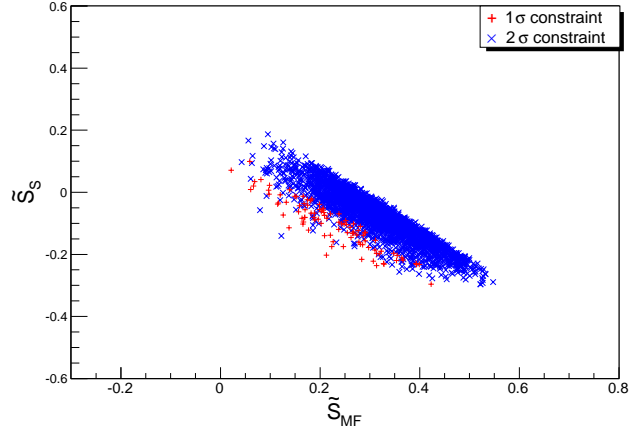} 
 \caption{{\small The plot shows the contribution to the S-parameter for the scalar sector ($\tilde{S}_S$) vs the mirror fermion sector ($\tilde{S}_{MF}$) within the 1 and 2 $\sigma$'s allowed region. The negative contribution to the S-parameter from the scalar sector tends to partially cancel the positive contribution from the mirror fermion sector and the total sum of the two contributions agrees with experimental constraints. }}
\label{SsvsSmf}
\end{figure}
In the above plot, \cite{hung2} took for illustrative purpose 3500 data points that fall inside the 2$\sigma$ region with about 100 points falling inside the 1$\sigma$ region. More details can be found in \cite{hung2}. 

\section {Review of the scalar sector of the EW $\nu_R$ model in light of the discovery of the 125-GeV SM-like scalar \cite{hung3}}

In light of the discovery of the 125-GeV SM-like scalar, it is imperative that any model beyond the SM (BSM) shows a scalar spectrum that contains at least one Higgs field with the desired properties as required by experiment. The present data from CMS and ATLAS only show signal strengths that are compatible with the SM Higgs boson. The definition of a signal strength $\mu$ is as follows 
\be
\sigma(H \text{-decay}) = \sigma(H \text{-production}) \times BR(H \text{-decay})\,,
\ee
and
\be\label{eq:mudef}
	\mu(H \text{-decay}) = \frac{\sigma(H \text{-decay})}{\sigma_{SM}(H \text{-decay})}\,.
\ee

To really distinguish the SM Higgs field from its impostor, it is necessary to measure the partial decay widths and the various branching ratios. In the present absence of such quantities, the best one can do is to present cases which are consistent with the experimental signal strengths. This is what was carried out in \cite{hung3}. 

The minimization of the potential containing the scalars shown above breaks its global symmetry $SU(2)_L \times SU(2)_R$ down to a custodial symmetry $SU(2)_D$ which guarantees at tree level $\rho = M_{W}^2/M_{Z}^2 \cos^2 \theta_W=1$ \cite{hung3}. The physical scalars can be grouped, based on their transformation properties under $SU(2)_D$ as follows:
	\begin{eqnarray}
		\text{five-plet (quintet)} &\rightarrow& H_5^{\pm\pm},\; H_5^\pm,\; H_5^0;\nonumber\\[0.5em]
		\text{triplet} &\rightarrow& H_{3}^\pm,\; H_{3}^0;\nonumber\\[0.5em]
		\text{triplet} &\rightarrow& H_{3M}^\pm,\; H_{3M}^0;\nonumber\\[0.5em]
		\text{three singlets} &\rightarrow& H_1^0,\; H_{1M}^0,\; H_1^{0\prime}\,,
	\end{eqnarray}
  The three custodial singlets are the CP-even states, one combination of which can be the 125-GeV scalar. In terms of the original fields, one has $H_1^0 = \phi_{2}^{0r}$,  $H_{1M}^0 = \phi_{2M}^{0r}$, and $H_1^{0\prime} = \frac{1}{\sqrt{3}} \Big(\sqrt{2}\chi^{0r}+ \xi^0\Big)$. These states mix through a mass matrix obtained from the potential and the mass eigenstates are denoted by $\widetilde{H}$, $\widetilde{H}^\prime$, and $\widetilde{H}^{\prime\prime}$, with the convention that the lightest of the three is denoted by $\widetilde{H}$, the next heavier one by $\widetilde{H}^\prime$ and the heaviest state by $\widetilde{H}^{\prime\prime}$. 
  
  To compute the signal strengths $\mu$, Ref.~\cite{hung3} considers $\widetilde{H} \rightarrow ZZ,~W^+W^-,~\gamma\gamma,~b\bar{b},~\tau\bar{\tau}$. In addition, the cross section of $g g \rightarrow \widetilde{H}$ related to $\widetilde{H} \rightarrow g g$ was also calculated. A scan over the parameter space of the model yielded {\em two interesting scenarios} for the 125-GeV scalar: 1) {\em Dr Jekyll}'s scenario in which $\widetilde{H} \sim H_1^0$ meaning that the SM-like component $H_1^0 = \phi_{2}^{0r}$ is {\em dominant}; 2) 
{\em Mr Hyde}'s scenario in which $\widetilde{H} \sim H_1^{0\prime}$ meaning that the SM-like component $H_1^0 = \phi_{2}^{0r}$ is {\em subdominant}. Both scenarios give signal strengths compatible with experimental data as shown below in Fig.~(2).
\begin{figure}
\label{signal2}
	\centering
	\includegraphics[width=0.5\textwidth]{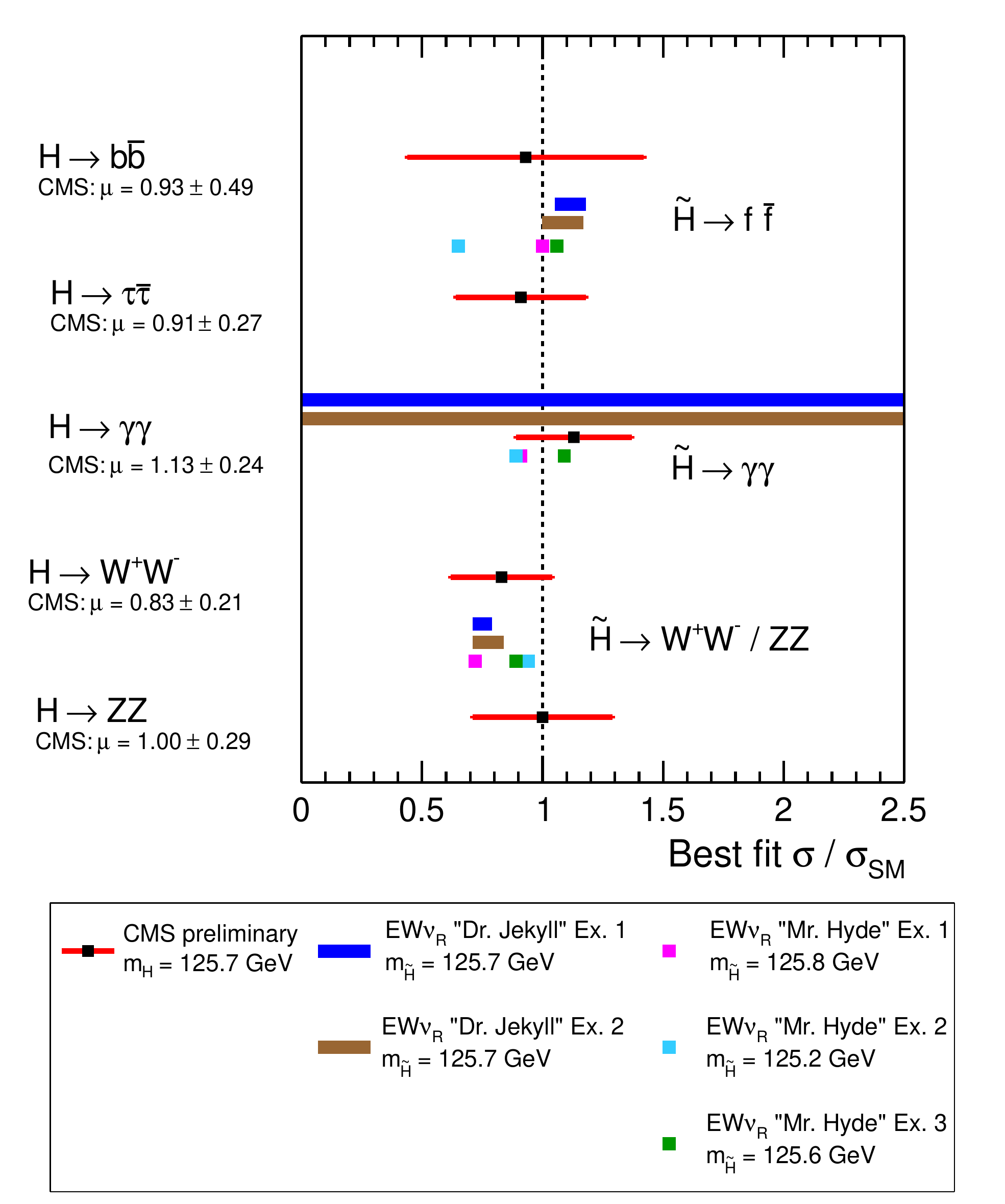}
	\caption{Figure shows the predictions of $\mu(\widetilde{H} \rightarrow ~b\bar{b}, ~\tau\bar{\tau}, ~\gamma\gamma, ~W^+W^-, ~ZZ)$ in the EW $\nu_R$ model for examples 1 and 2 in {\em Dr.~Jekyll} and example 1, 2 and 3 in {\em Mr.~Hyde} scenarios as discussed in \cite{hung3}, in comparison with corresponding best fit values by CMS \cite{h_ww_122013, h_zz_4l_122013, h_bb_102013, h_tautau_012014}.}
\end{figure}

As we can see from Fig.~(2), both SM-like scenario ({\em Dr Jekyll}) and the {\em more interesting scenario} which is very unlike the SM ({\em Mr Hyde}) agree with experiment. As stressed in \cite{hung3}, present data cannot tell whether or not the 125-GeV scalar is truly SM-like or even if it has a dominant SM-like component. It has also been stressed in \cite{hung3} that it is essential to measure the partial decay widths of the 125-GeV scalar to truly reveal its nature. Last but not least, in both scenarios, $H_{1M}^0 = \phi_{2M}^{0r}$ is subdominant but is essential to obtain the agreement with the data as shown in \cite{hung3}.

As discussed in detail in \cite{hung3} , for proper vacuum alignment, the potential contains a term proportional to $\lambda_5$ (Eq.~(32) of \cite{hung3}) and it is this term that prevents the appearance of Nambu-Goldstone (NG) bosons in the model. The would-be NG bosons acquire a mass proportional to $\lambda_5$ .

An analysis of CP-odd scalar states $H_{3}^0, H_{3M}^0 $ and the heavy CP-even states $\widetilde{H}^\prime$, and $\widetilde{H}^{\prime\prime}$ was presented in \cite{hung3}. The phenomenology of charged scalars including the doubly-charged ones was also discussed in \cite{pqaranda}.

The phenomenology of mirror quarks and leptons was briefly discussed in \cite{Hung} and a detailed analysis of mirror quarks was presented in \cite{CGHHN}. It suffices to mention here that mirror fermions decay into SM fermions through the process $q^M\rightarrow q\phi_S$, $l^M\rightarrow l\phi_S$ with $\phi_S$ "appearing" as missing energy in the detector. Furthermore, the decay of mirror fermions into SM ones can happen outside the beam pipe and inside the silicon vertex detector. Searches for non-SM fermions do not apply in this case. It is beyond the scope of the paper to discuss these details here.

\section{Yukawa interactions between mirror and SM leptons}



The EW $\nu_R$ model has been extended to include an investigation of neutrino and charged lepton mass matrices and mixings \cite{pqtrinh}. In \cite{pqtrinh}, a non-abelian discrete symmetry group $A_4$ was assumed and was applied to the Higgs singlet sector which is responsible for the Dirac masses of the neutrinos. Following \cite{pqtrinh}, we list the assignments of the SM and mirror fermions as well as those for the scalars under $A_4$.
\begin{table}[!htb]
\caption{\label{assignment} $A_4$ assignments for leptons and Higgs fields}
\begin{center}
\begin{tabular}{| c || c | c | c | c | c | c | c |}
 \hline
 Field & $\mathnormal{(\nu,l)_L}$ & $\mathnormal{(\nu, l^M)_R}$ & $\mathnormal{e_R}$ & $\mathnormal{e_L^M}$ & $\mathnormal{\phi_{0S}}$ & $\mathnormal{\tilde{\phi}_{S}}$ & $\mathnormal{\Phi_2}$ \\ [0.5ex]
 \hline
 $A_4$ & $\underline{3}$ & \underline{3} & \underline{3} & \underline{3} & \underline{1} & \underline{3} & \underline{1}\\
  \hline
\end{tabular}
\end{center}
\end{table} 
From this assignment, one obtains the following Yukawa interactions in terms of lepton mass eigenstates ($l^{e}_L=(e_L,\mu_L,\tau_L)$, $l^{{\nu}_e}_L=({\nu_e}_L,{\nu_\mu}_L,{\nu_\tau}_L)$, $l^{M,e}_{R}=(e^{M}_R, \mu^{M}_R, \tau^{M}_R)$, $l^{M,{\nu}_e}_{R}=({\nu_e}^{M}_R, {\nu_\mu}^{M}_R, {\nu_\tau}^{M}_R)$):
\bea
\label{eq:yukawadown}
L_S&=& \bar{l}^{e}_{L}\, U^{e\dagger}_{L}  M^{e}_{\phi}  U^{e^M}_R \, l^{M,e}_{R} +H.c. \; \; \nonumber \\
        &=& \bar{l}^{e}_{L}\, \bar{M^{e}_{\phi}}  \, l^{M,e}_{R} +H.c. \; 
\eea
for the electron sector and
\bea
\label{eq:yukawaup}
L_S&=& \bar{l}^{{\nu_e}}_{L}\, U^{{\nu_e}\dagger}_{L}  M^{{\nu_e}}_{\phi}  U^{{\nu_e}^M}_R \, l^{M,{\nu_e}}_{R} +H.c. \; \; \nonumber \\
        &=& \bar{l}^{{\nu_e}}_{L}\, \bar{M^{{\nu_e}}_{\phi}}  \, l^{M,{\nu_e}}_{R} +H.c. \; 
\eea
for the electron neutrino sector and where
\be
\label{eq:mnu}
M^{e,{\nu_e}}_{\phi} = 
\left(
  \begin{array}{cccc}
    g^{e,{\nu_e}}_{0S}\phi_{0S} & g^{e,{\nu_e}}_{1S}\phi_{3S} & g^{e,{\nu_e}}_{2S}\phi_{2S} \\
    g^{e,{\nu_e}}_{2S}\phi_{3S} & g^{e,{\nu_e}}_{0S}\phi_{0S}  & g^{e,{\nu_e}}_{1S}\phi_{1S} \\
    g^{e,{\nu_e}}_{1S}\phi_{2S}  & g^{e,{\nu_e}}_{2S}\phi_{1S} & g^{e,{\nu_e}}_{0S}\phi_{0S}  \\
  \end{array}
\right) \, .
\ee

The mixing parameters involving in the decay $l^{M,i}_R \rightarrow l^{j}_L + \phi_l$ where $i$ and $j$ denote quark flavors and $l=0,..,3$ are contained in the parametrization of $\bar{M}^{e,{\nu_e}}_{\phi}$ as well as in Eq.~(\ref{eq:mnu}).

An important remark is in order here. The Yukawa couplings $g_{Sl}$ in the lepton sector are constrained by rare processes such as $\mu \rightarrow e \, \gamma$ and has been studied in detail in this previous work. \cite{Hung:2015hra}


As shown in Eqs: (\ref{eq:yukawadown}, \ref{eq:yukawaup}, \ref{eq:mnu}), the right-handed neutrinos and the mirror charged leptons couple to the SM counterparts through the scalar singlets $\phi_S$. The decay width of the process such as $l^M\rightarrow l+\phi_S$ depends on the coupling $g_{Sl}$. In general, 
\bea
\Gamma(l^M\rightarrow l+\phi^{\star}_S)~=~\dfrac{g_{Sl}^2}{64\pi}m_{l^M}\left(1-\frac{m_l^2}{m_{l^M}^2}\right)\left(1+\frac{m_l}{m_{l^M}}-\frac{m_l^2}{2m_{l^M}^2}\right)
\eea

\section{ Phenomenology: Collider signals at the LHC}

In this section, we will discuss the collider signatures of mirror leptons (charged mirror leptons, $e^{M\pm}$, as well as right handed mirror neutrinos, $\nu_R^M$, in the framework of {\em EW $\nu_R$} model. Since the masses of mirror leptons being restricted to be in the ballpark of few hundred GeVs from the perturbativity of the Yukawa couplings, the pair production cross-section of the mirror leptons could be significant enough to probe or ruled out the {\em EW $\nu_R$} model at the ongoing/future run of the LHC. Therefore, it is instructive to study the collider signatures of mirror leptons in the framework of {\em EW $\nu_R$} model. Mirror leptons have gauge coupling with photon, $W^\pm$ and $Z$-boson. Therefore, pair-productions of mirror leptons at the LHC take place through quark antiquark initiated processes with a $\gamma/W^\pm/Z$ in the s-channel. For example, the pair production of charged mirror leptons, $\bar e^M e^M$, (right handed mirror neutrinos, $\nu^M_R\nu^M_R$) proceeds via a photon or $Z$-boson (only $Z$-boson) exchange in the s-channel, whereas, $e^M \nu_R^M$ production takes place via $W^\pm$ exchange. After being produced, the mirror leptons decay into SM quarks, leptons, neutrinos and the singlet scalar, $\phi_S$. The final state neutrinos and $\phi_S$ remain elusive at the detector and thus, give rise to missing energy signature. Before going into the detailed discussion of collider signature, it is important to discuss the decay modes of the mirror leptons. Assuming right handed mirror neutrino ($\nu_R^M$) being heavier than the charged mirror lepton ($e^M$), there are two possible decay modes for the $\nu_R^M$. It can decay into a SM neutrino ($\nu_L$) and $\phi_S$. This decay takes place via the Yukawa interaction in Eq.~\ref{dirac} and hence, suppressed due to Yukawa coupling which is required to be small ($<10^{-3}$) from the constraint coming from the $\mu \rightarrow e \gamma$ decay. $\nu_R^M$ dominantly decays into a $e^{M\pm}$ in association with a on/off shell (depending on the $e^M$--$\nu_R^M$ mass splitting)  $W^\mp$ which subsequent decays into a pair of jets or lepton-neutrino pair. The decay of $e^{M\pm}$ into a $W^\pm$ and $\nu_R^M$ is kinematically forbidden for $m_{e^M}<m_{\nu_R^M}$. Therefore, $e^{M\pm}$ decays into $e^\pm$ and $\phi_S$ with 100\% branching ratio. The resulting collider signatures of the production of $e^{M+}e^{M-}$, $e^{M\pm}_R\nu_R^M$ and $\bar \nu_R^M \nu_R^M$ are summarized in the following:\\
1. $e^{M+}e^{M-}$-pair production gives rise to opposite sign dilepton (OSD) in association with missing transverse energy signature at the collider: 
\be 
pp \to e^{M+}e^{M-} \to (e^+ \phi_S)(e^-\phi_S) \to e^+e^- + p_T\!\!\!\!\!\!/~. \nonumber
\ee
2. Pair production of $e^{M\pm}_R\nu_R^M$ gives rise to 2-lepton (opposite or same sign) and 3-lepton signatures at the collider. After being produced, $e^{M\pm}_R$ decays into a SM charged lepton ($e^\pm$) and $\phi_S$. Whereas, $\nu_R^M$, being heavier than $e^{M\pm}$, decays into a $e^\pm$ in association with a on/off shell (depending on the $m_{e^M}$--$m_{\nu_R^M}$ mass splitting) $W^\mp$, which finally decays to a pair of jets or charged-lepton + neutrino pairs giving rise to same sign dilepton (SSD) and OSD or trilepton + missing transverse energy signatures, respectively. Missing transverse energy results from the elusive $\phi_S$ and neutrinos in the final state. The SSD and OSD signatures are always accompanied by a pair of jets arising from the on/off shell $W$-decay. Whereas, at the parton-level, tri-lepton signature is not accompanied by any hadronic activity.
\begin{table}[!htb]
\begin{center}
\begin{tabular}{ccccccc}
     &               &                   &                  &                                 & &$e^\pm e^\pm q q^\prime \phi_S \phi_S$ (SSD+2-jets + $p_T\!\!\!\!\!\!/~$)\\
     &               &                   &                  & $(e^\pm \phi_S)(e_R^{M\pm} W^\mp)$ & $^\nearrow_\searrow$ &                                    \\
$pp$ & $\rightarrow$ & $e^{M\pm}_R\nu_R^M$ & $^\nearrow_\searrow$ &                                 & &$e^\pm e^\pm e^\mp \nu_L \phi_S \phi_S$ (3-leptons + $p_T\!\!\!\!\!\!/~$)\\ 
     &               &                   &                   & $(e^\pm \phi_S)(e_R^{M\mp} W^\pm)$ & $^\nearrow_\searrow$ &                       \\
     &               &                   &                   &                                 &   & $e^\pm e^\mp q q^\prime \phi_S \phi_S$ (OSD+2-jets + $p_T\!\!\!\!\!\!/~$)\\
\end{tabular}
\end{center}
\end{table} 
\\
3. Similarly, pair production of $\nu_R^M \nu_R^M$ gives rise to 2-leptons (OSD and SSD), 3-leptons as well as 4-leptons in association with jets and missing transverse momentum signatures. 
\begin{table}[!htb]
\begin{center}
\begin{tabular}{ccccccc}
     &               &                   &                  &                                 & &$e^\pm e^\pm q q^\prime q q^\prime \phi_S \phi_S$ (SSD+4-jets + $p_T\!\!\!\!\!\!/~$)\\
     &               &                   &                  & $(e_R^{M\pm} W^\mp)(e_R^{M\pm} W^\mp)$ & $^\nearrow_\searrow \!\!\!\!\!\!\rightarrow$ & $e^\pm e^\pm e^\mp \nu_L q q^\prime \phi_S \phi_S$ (3l+2-jets + $p_T\!\!\!\!\!\!/~$)                                    \\
$pp$ & $\rightarrow$ & $\nu_R^{M}\nu_R^M$ & $^\nearrow_\searrow$ &                                 & &$e^\pm e^\pm e^\mp e^\mp \nu_L \nu_L\phi_S \phi_S$ (4-leptons + $p_T\!\!\!\!\!\!/~$)\\ 
     &               &                   &                   & $(e_R^{M\pm} W^\mp)(e_R^{M\mp} W^\pm)$ & $^\nearrow_\searrow \!\!\!\!\!\!\rightarrow$ & $e^\pm e^\pm e^\mp \nu_L q q^\prime \phi_S \phi_S$ (3l+2-jets + $p_T\!\!\!\!\!\!/~$)                      \\
     &               &                   &                   &                                 &   & $e^\pm e^\mp q q^\prime q q^\prime \phi_S \phi_S$ (OSD+4-jets + $p_T\!\!\!\!\!\!/~$)\\
\end{tabular}
\end{center}
\end{table} 
OSD + $p_T\!\!\!\!\!\!/~$ signal suffers from the usual drawback of dealing with huge SM background contributions arising mainly from $W^+W^-$ and $\bar t t$ production. On the other hand, 3-lepton and 4-lepton signals are suppressed by the leptonic branching ratio of the $W$-boson. Therefore, in this work, we have studied same sign dilepton (SSD) in association with jets and $p_T\!\!\!\!\!\!/~$ as a signature of {\em EW $\nu_R$} model. 

\begin{figure}[t]
\includegraphics[scale=0.26]{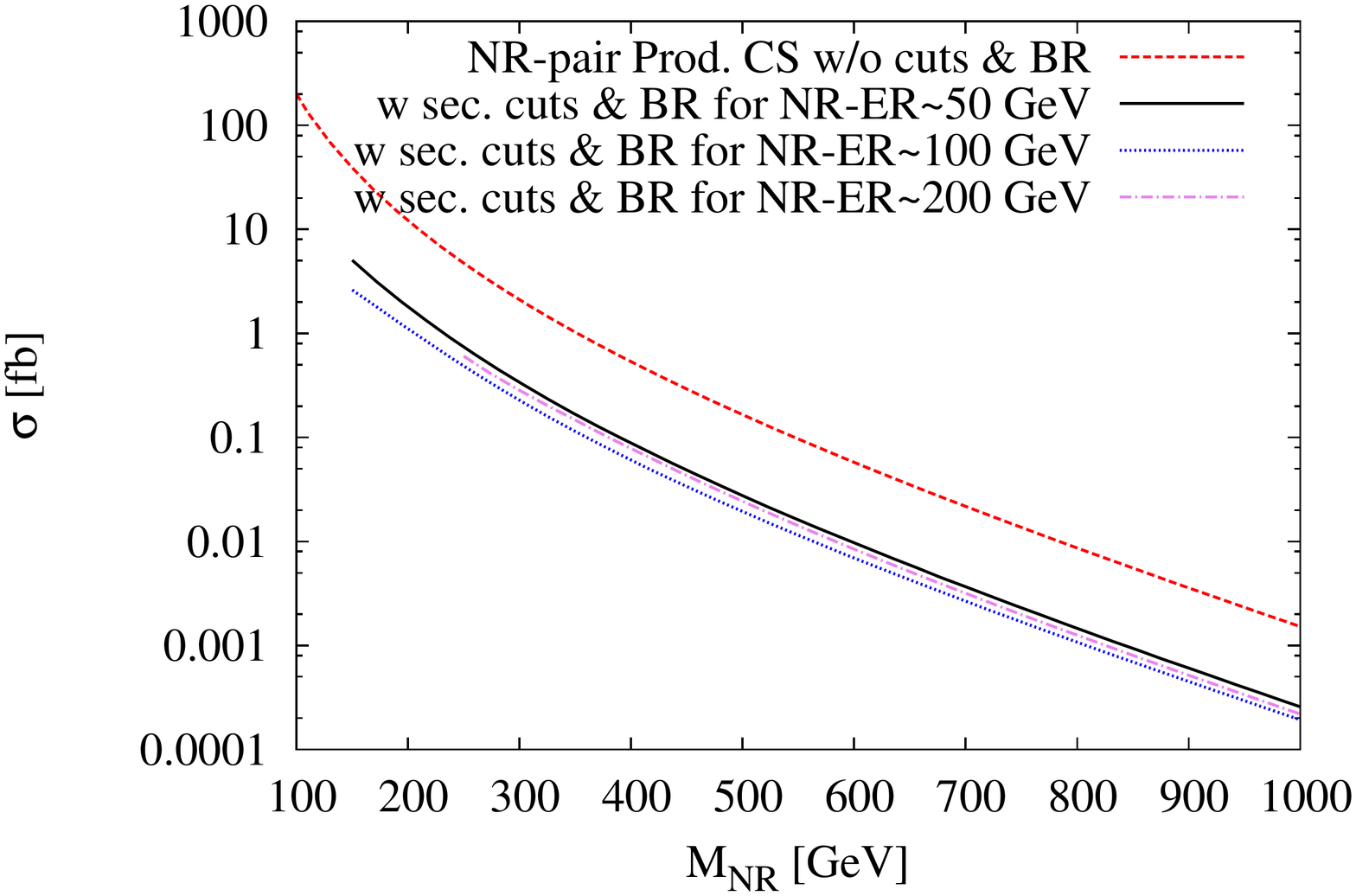} 
\includegraphics[scale=0.26]{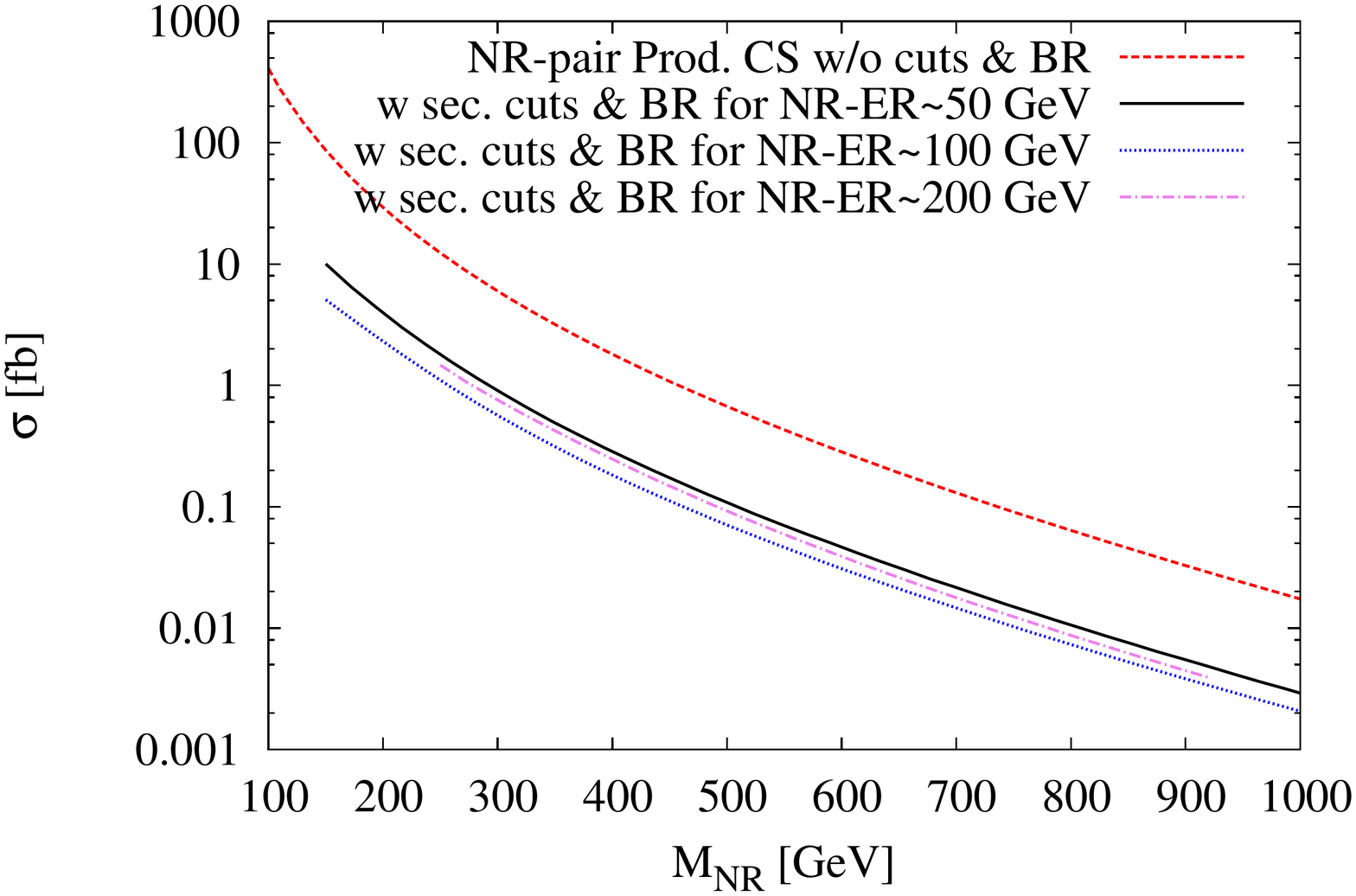} 
\caption{$\nu_R^M$-pair production cross-sections as a function of $\nu^M_R$-mass ($M_{NR}$) for 8 TeV (left panel) and 13 TeV (right panel) center of mass energy of the LHC. We have also presented resulting SSD signal cross-sections after multiplying with the branching ratios and imposing the acceptance cuts (listed in Eqs. \ref{cut:pT}--\ref{cut:jj-iso}) for different values of $e^M$--$\nu_R^M$ mass splitting.}
\label{cross_NRNR}
\end{figure}

As discussed in the previous paragraph, SSD signature arises from the production and decay of $e^{M\pm}_R\nu_R^M$ and $\nu_R^M \nu_R^M$ pairs. The spin averaged matrix element squared for the above mentioned productions are given by,
\begin{flalign}
\label{eq:Msquared1}
|M(q\bar q \to \bar \nu_R^M \nu_R^M)|^2=&\dfrac{g^4}{24c_W^4}\dfrac{1}{(\hat s-M_Z^2)^2+\Gamma_Z^2M_Z^2}\nonumber\\
            \times&\big{[}(A_q^2+V_q^2)(A_{\nu}^2+V_{\nu}^2)\left((M_R^2-t)^2+(M_R^2-u)^2 \right)\nonumber\\
                +&M_R^2(A_{\nu}^2-V_{\nu}^2)(A_q^2+V_q^2)\dfrac{s}{2}-4A_qV_qA_{\nu}V_{\nu}\left((M_R^2-t)^2 -(M_R^2-u)^2\right)\big{]},
 \end{flalign}
\begin{flalign}                
\label{eq:Msquared2}  
|M(q\bar q^\prime \to \bar e^M \nu_R^M)|^2=&\dfrac{g^4}{12}\frac{1}{[s-M_W^2]^2+\Gamma_W^2M_W^2}(m_{l^M}^2-t)(M_R^2-t),
\end{flalign}
where $s,t$ and $u$ are Mandelstam variables. In the Eq.~(\ref{eq:Msquared1}), $A_q,~V_q$ are the axial and vector couplings in the $\bar{q}qZ$ interaction \cite{PDG2014}
\begin{flalign}
\label{eq:qqZcurrent}
\mathcal{L}_{\bar{q}qZ}~=~-\dfrac{g}{2\cos\theta_W}\bar{q}\gamma^{\mu}(V_q-A_q\gamma^5)qZ_{\mu}.
\end{flalign}
 While $A_{\nu},~V_{\nu}$ are the axial and vector couplings in the $\nu_R\nu_RZ$ interaction \cite{hung2}
\begin{flalign}
\label{eq:nunuZcurrent}
\mathcal{L}_{\nu_R\nu_RZ}~=~-\dfrac{g}{2\cos\theta_W}\nu_R\gamma^{\mu}(V_{\nu}-A_{\nu}\gamma^5)\nu_RZ_{\mu}.
\end{flalign}
Since the right-handed neutrinos ($\nu_R$) in the $EW\nu_R$ model have Majorana nature, so that $A_{\nu}=1,\;V_{\nu}=0$. \cite{MS}

\begin{table}[ht]
\caption{Couplings of fermions, including the right-handed neutrino $\nu_R$ to the Z-boson. Here, $\theta_W$ is the electroweak mixing angle ($sin^2\theta_W \simeq 0.231$)}
\centering
\begin{tabular}{c c c}
Fermions & $V_f$ & $A_f$\\
\hline
u, c, t   & $+1/2 - 4/3sin^2\theta_W$ & +1/2\\
d, s, b & $-1/2 +2/3sin^2\theta_W$ & -1/2\\
$\nu_R$ & 0 & 1\\
\hline
\end{tabular}
\label{table:nonlin}
\end{table}


The pair production cross-sections at the LHC are obtained by integrating the spin averaged matrix element squared over the phase-space and parton distribution functions.  For numerical evaluation of the cross-sections, we use a tree-level Monte-Carlo program incorporating CTEQ6L1 \cite{cteq} parton distribution functions. Both the renormalization and the factorization scales have been set equal to the subprocess center-of-mass energy $\sqrt{\hat s}$. 

\begin{figure}[t]
\includegraphics[scale=1]{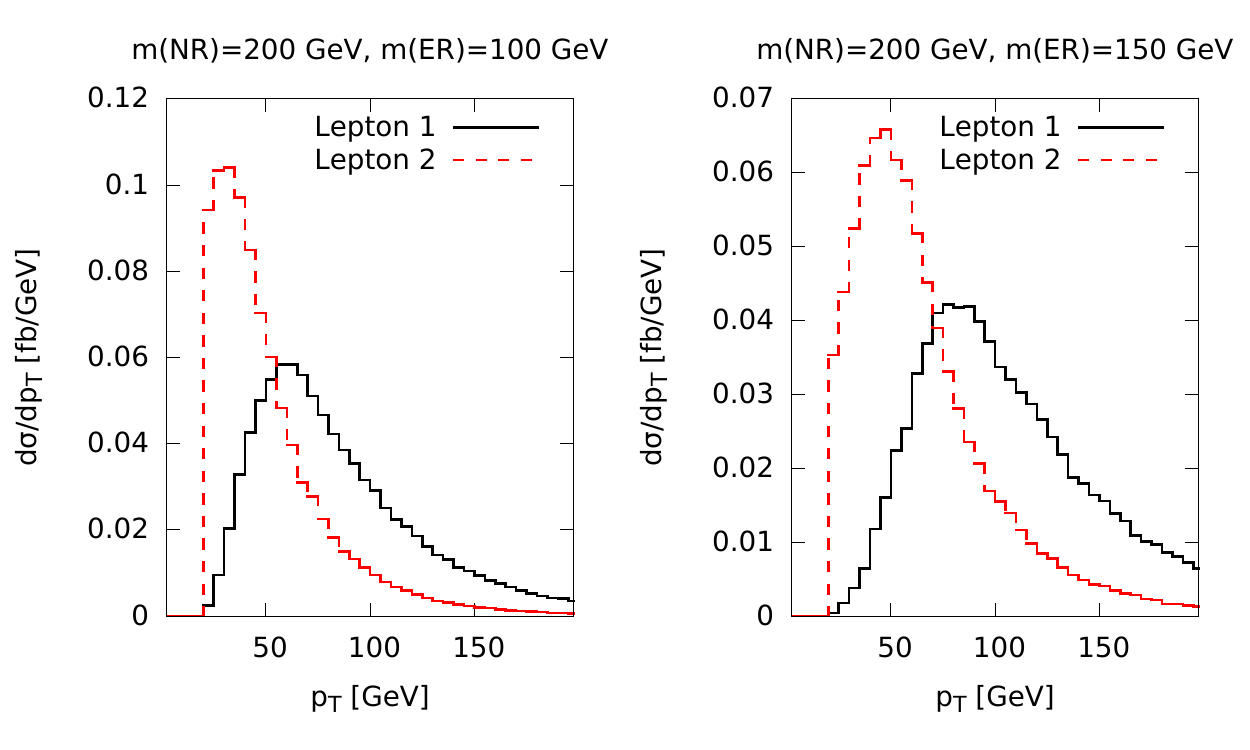} \\
\includegraphics[scale=1]{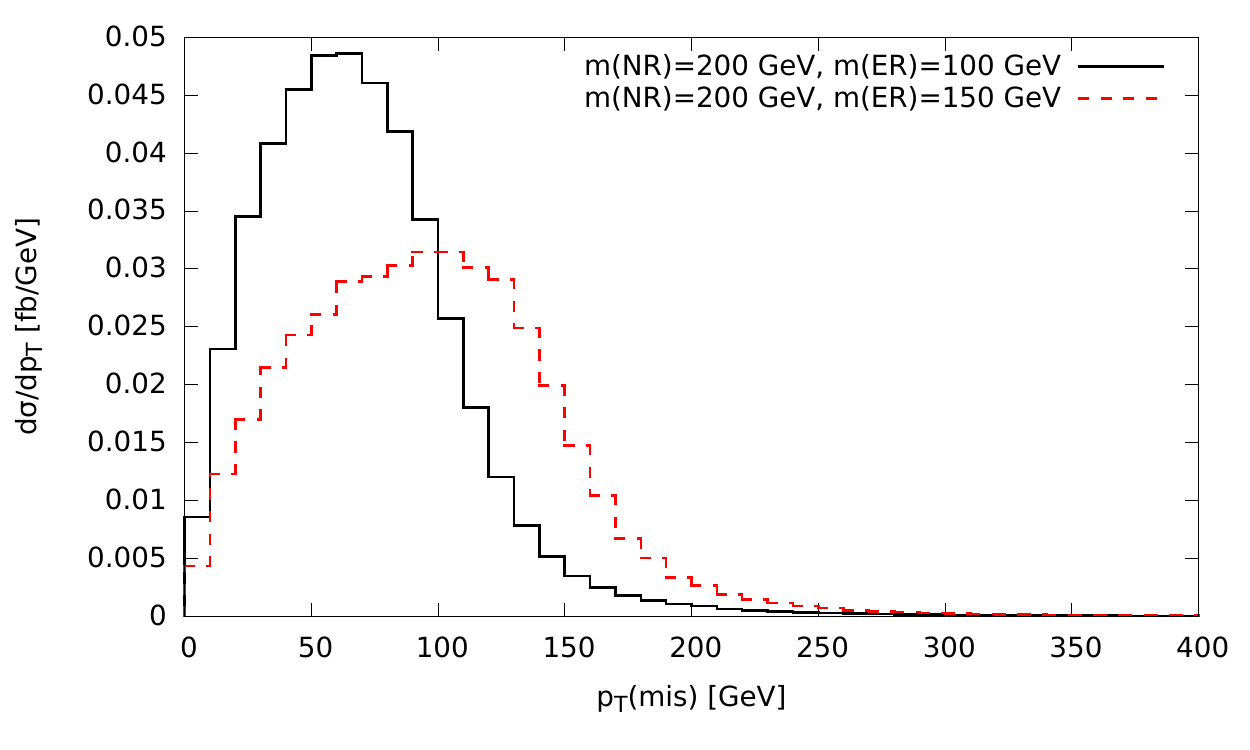} 
\caption{Transverse momentum distributions (top panel) and missing $p_T$ distributions (bottom panel) of same-sign dileptons resulting from $\nu^M_R \nu^M_R$ production at the LHC with $\sqrt s=13$ TeV for two different values of $e^M$--$\nu_R^M$ mass splitting. The distributions are plotted with the leptons after ordering them according to their $p_T$ hardness ($p_T^{l_1}>p_T^{l_2}$). }
\label{dist_NRNR}
\end{figure}

At this stage, we are equipped enough to compute SSD signal cross-section as well as characteristic kinematic distributions. However, before going into the discussion of cross-section and distributions, it is important to list a set of basic requirements for leptons and jets to be visible at the detector. It should be noted that any realistic detector has only a finite resolution; this applies to both energy/transverse momentum measurements as well as the determination of the angle of motion. For our purpose, the latter effect can be safely neglected
\footnote{The angular resolution is, generically, far
superior to the energy/momentum resolutions and too fine to be of any
consequence at the level of sophistication of this analysis.} 
and we simulate the former by smearing the energy with Gaussian
functions. The energy resolution function receives contributions from
many sources and are, in general, a function of the detector
coordinates. We, though, choose to simplify the task by assuming a
flat resolution function equating it to the worst applicable for our
range of interest \cite{gsmear}, namely, 
\be
\frac{\Delta E}{E}=\frac{a}{\sqrt {E/{\rm GeV}}}\oplus b,
\ee
where,  $ a=100\%, b=5\%$ for jets and $a=15\%$ and $b=1\%$ for leptons, and $\oplus $ denotes a sum in quadrature. Keeping in mind the LHC environment as well as the detector configurations, 
we demand that, to be visible, a lepton must have an 
adequately large transverse momentum and they are well inside 
the rapidity coverage of the detector, 
namely,
\be
p_T^{l} > 20~GeV \ ,
\label{cut:pT}
\ee
\be
|\eta_{l}| \leq 2.5 \ .
\label{cut:eta}
\ee
We demand that a lepton be well separated from other leptons and jets so that they can be identified as individual entities. To this end, we use the well-known cone algorithm defined in terms of a cone angle $\Delta R_{ij} \equiv \sqrt{\left (\Delta \phi_{ij}\right)^2 
+ \left(\Delta \eta_{ij}\right)^2} $, with 
$\Delta \phi $ and $ \Delta \eta $ being the azimuthal angular 
separation and rapidity difference between two particles.
Quantitatively, we impose
\be
\Delta R_{l \, l} > 0.4.;~~~~~~
\Delta R_{l \, j} > 0.4.
\label{cut:jj-iso}
\ee
The requirements summarized  in  Eqs. (\ref{cut:pT}--\ref{cut:jj-iso}) constitute our {\em acceptance cuts}.

In Fig.~\ref{cross_NRNR}, we have presented production cross-sections of a pair of right handed mirror neutrinos as a function of its mass at the LHC with 8 TeV (left panel) and 13 TeV (right panel) center of mass energy. $\sigma({\nu_R^M \nu_R^M})$ varies from few 100 fb to about $10^{-3}$ fb as we vary $m_{\nu_R^M}$ between 100 GeV to 1 TeV. The signal under consideration is required to have two same sign charged leptons. In Fig.~\ref{cross_NRNR}, we have also presented SSD signal cross-sections after the acceptance cuts summarized in Eqs. (\ref{cut:pT}--\ref{cut:jj-iso}) for three different values of $e^M$--$\nu_R^M$ mass splitting namely, 50 GeV\footnote{In this case, $\nu_R^M$ decays to $e^M q \bar q^\prime$ via tree level 3-body decay involving a off shell $W$-boson.}, 100 GeV and 200 GeV\footnote{For this mass splitting, the decay of $\nu_R^M$ into a charged mirror lepton and $W$-boson is kinematically allowed. Therefore, $\nu_R^M$ decays into $e^{M\pm}W^{\mp}$ pairs followed by the decay of $W$-boson in to a pair of jets.}. The resulting transverse momentum distributions of leptons after ordering them according to their $p_T$ hardness ($p_T^{l_1}>p_T^{l_2}$) are presented in Fig.~\ref{dist_NRNR} (left two panels) for 50 GeV and 100 GeV splitting between $\nu_R^M$ and $e^M$. It is important to note that the leptons are arising from the decay $e^M \to e \phi_S$ and hence, will usually carry significant momentum due to relatively large mass  splitting between $e^M$ and the light singlet scalar $\phi_S$. The singlet scalars remain invisible in the detector and thus, give rise to an imbalance in the transverse momentum of the system known as missing transverse momentum.  The missing transverse momentum defined in terms of the 
total visible momentum, as,
\be
\not p_T \equiv \sqrt{ \bigg(\sum_{\rm vis.} p_x \bigg)^2 
                 + \bigg(\sum_{\rm vis.} p_y \bigg)^2 }.\nonumber
\ee 
In Fig.~\ref{dist_NRNR} (right panel), we have presented $\not p_T$ distributions associated with a pair of same sign leptons resulting from the production of $\nu_R^M \nu_R^M$ pairs. 

\begin{figure}[t]
\includegraphics[scale=0.26]{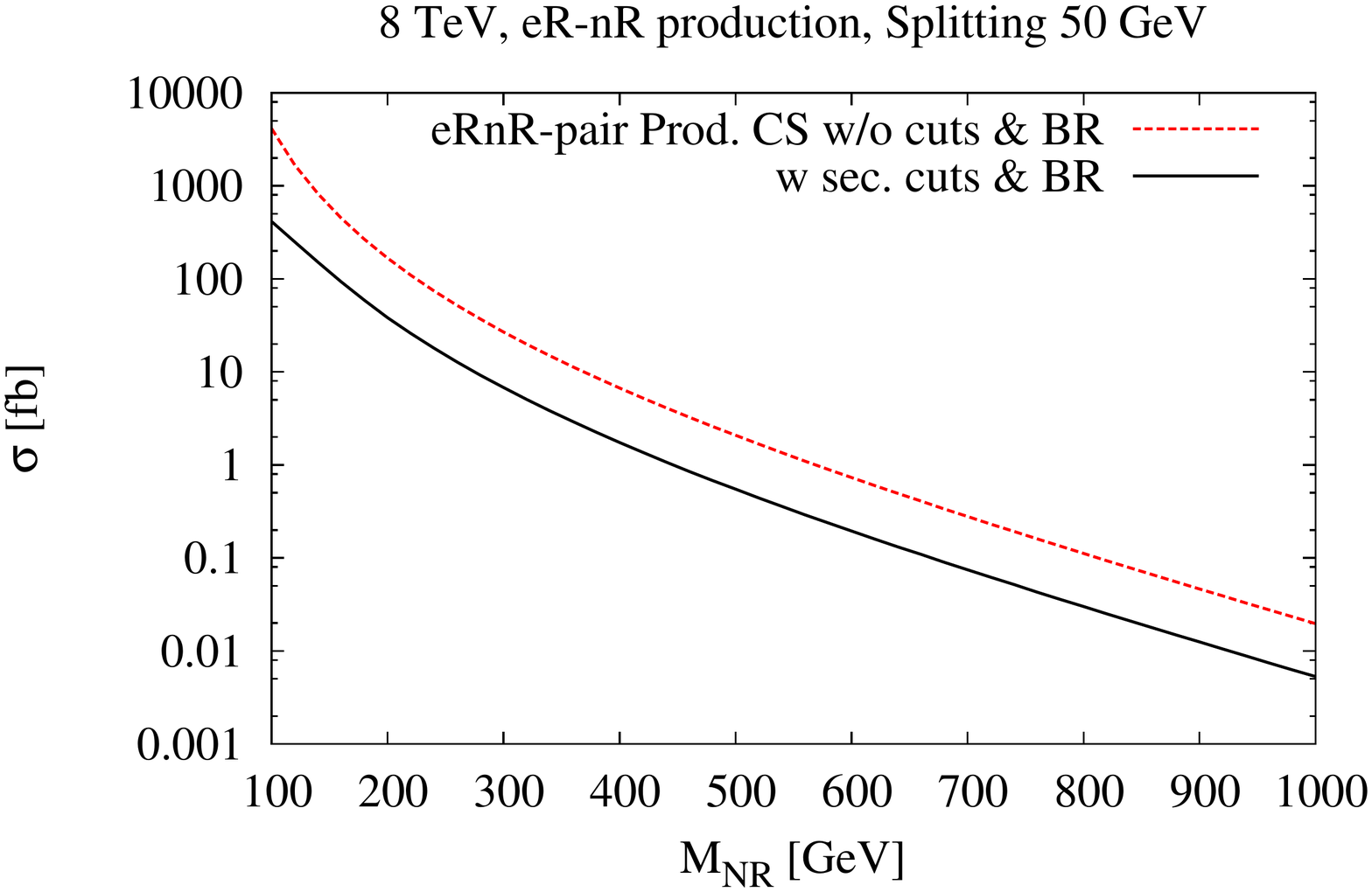} 
\includegraphics[scale=0.26]{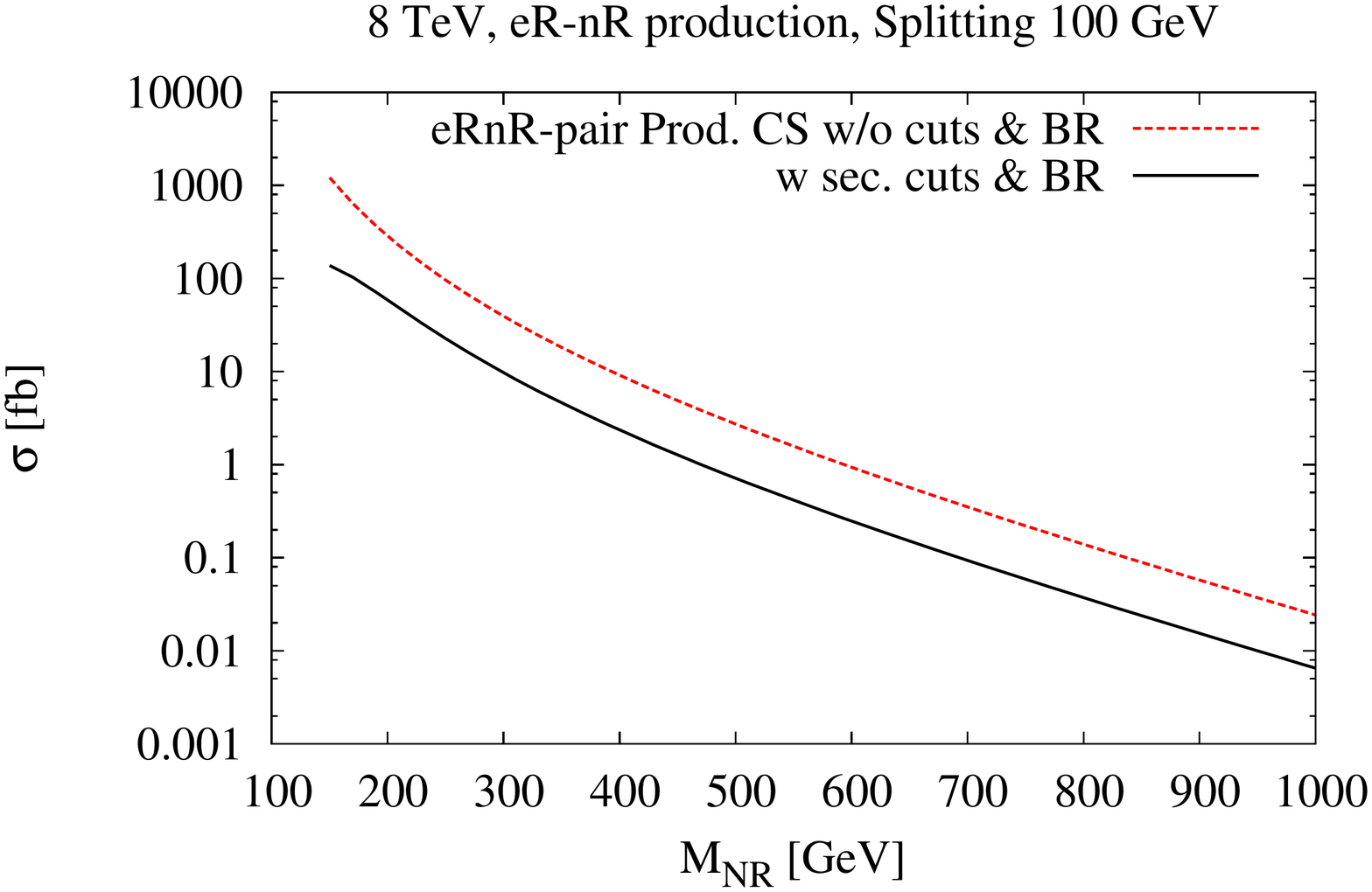} \\
\includegraphics[scale=0.26]{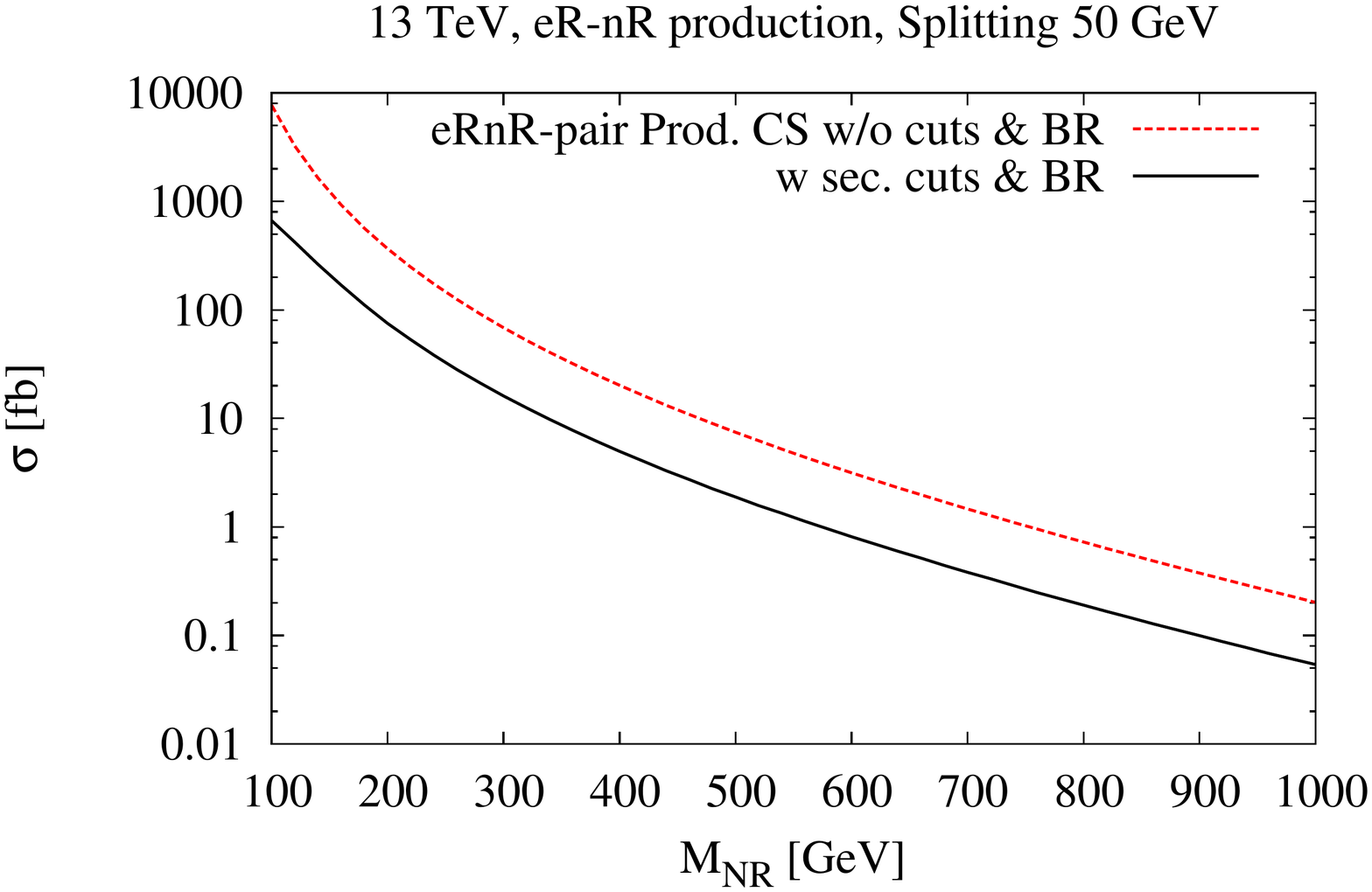} 
\includegraphics[scale=0.26]{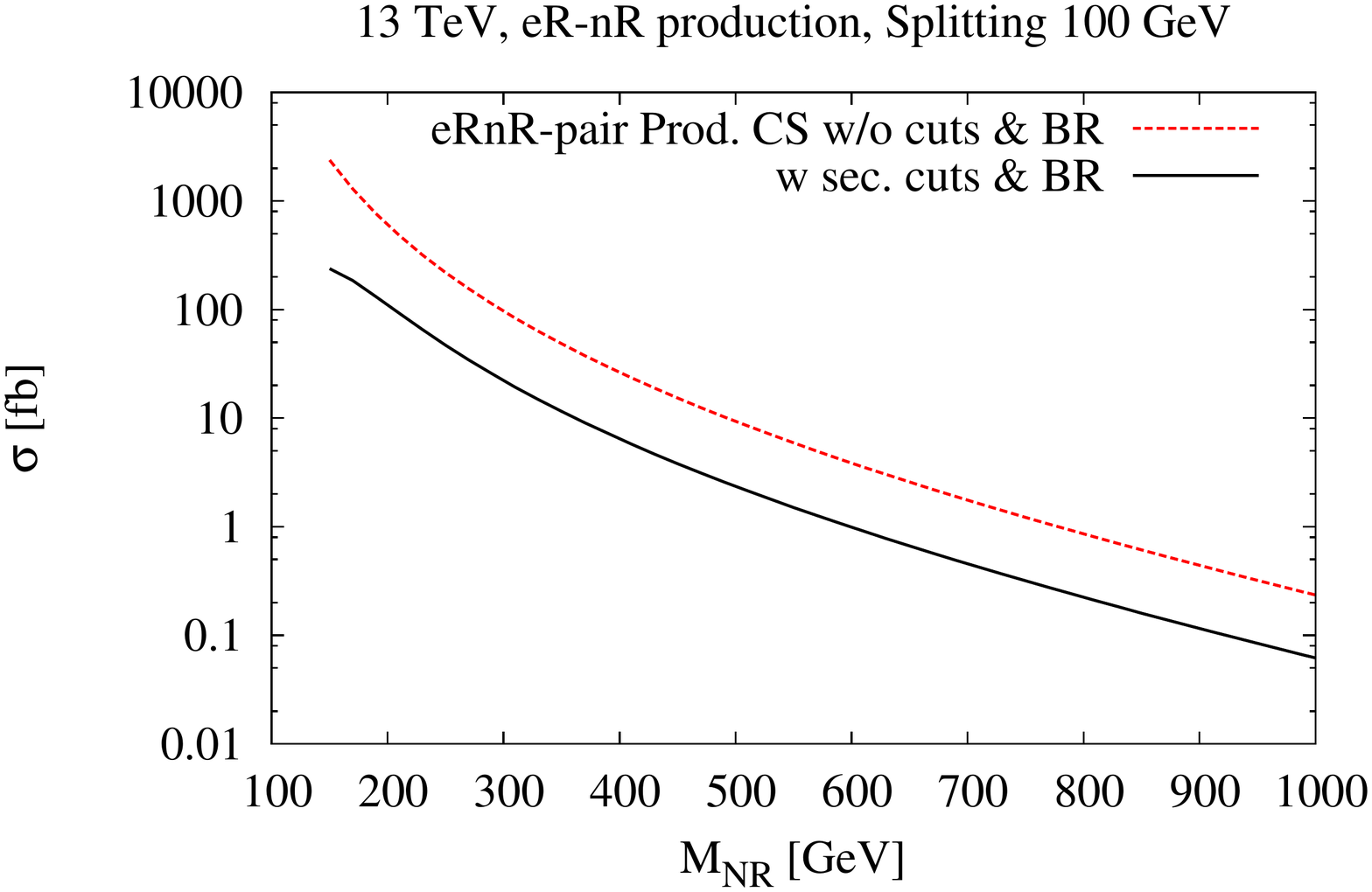} \\
\caption{$\nu_R^M e^M$ production cross-sections as a function of $\nu^M_R$-mass for 8 TeV (top panel) and 13 TeV (bottom panel) center of mass energy of the LHC and two different values, namely 50 GeV (left panel) and 100 GeV (right panel), of $e^M$--$\nu_R^M$ mass splitting. We have also presented resulting SSD signal cross-sections after multiplying the pair production cross-sections with the branching ratios and imposing the acceptance cuts (listed in Eqs. \ref{cut:pT}--\ref{cut:jj-iso}).}
\label{cross_ERNR}
\end{figure}

\begin{figure}[t]
\includegraphics[scale=1]{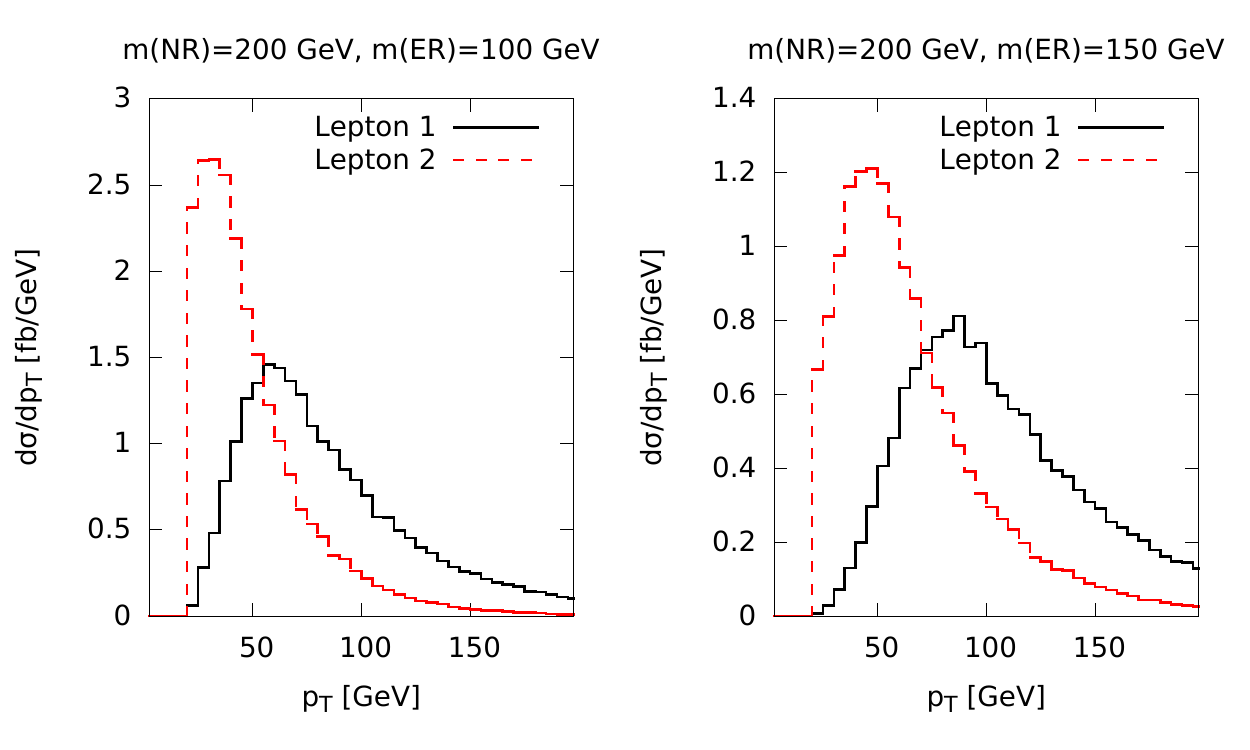}\\
\includegraphics[scale=1]{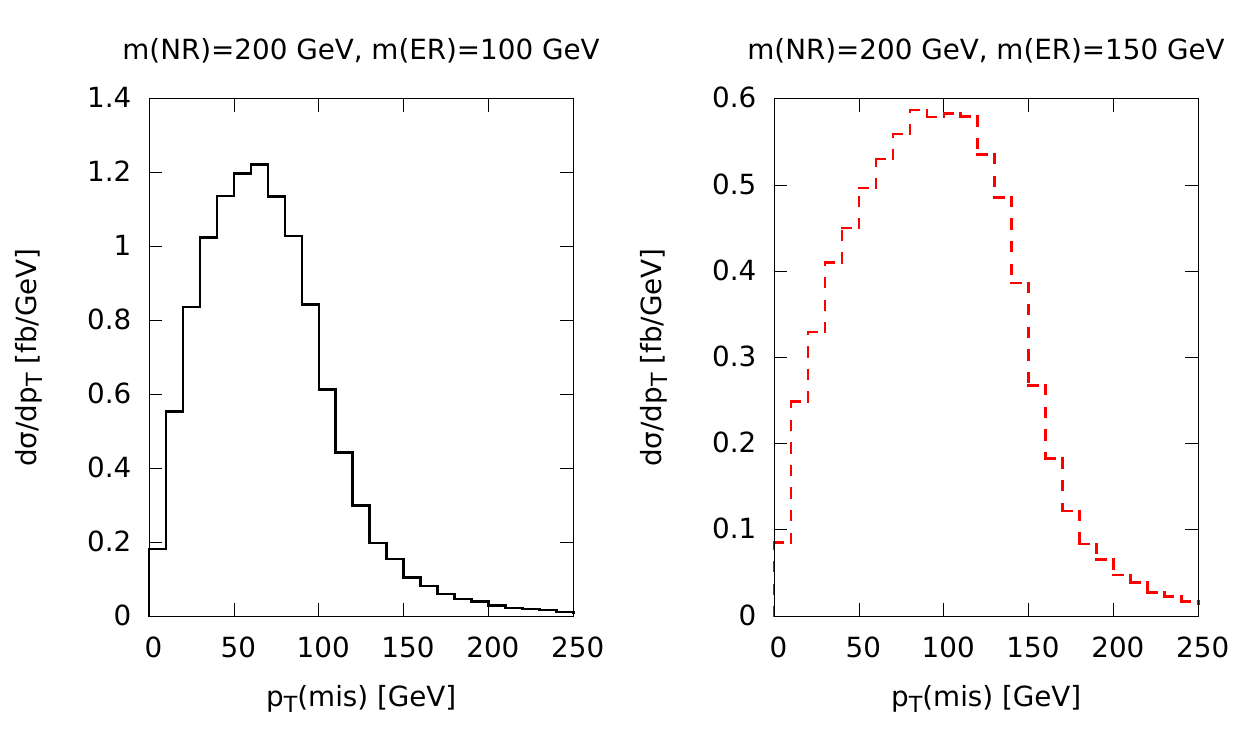} 
\caption{Transverse momentum distributions (top panel) and missing $p_T$ distributions (bottom panel) of same-sign dileptons resulting from $e^M \nu^M_R$ production at the LHC with $\sqrt s=13$ TeV for two different values of $e^M$--$\nu_R^M$ mass splitting. Leptons are ordered according to their $p_T$ hardness ($p_T^{l_1}>p_T^{l_2}$). }
\label{dist_ERNR}
\end{figure}

In Fig.~\ref{cross_ERNR}, we have presented $\sigma(\bar e^M \nu_R^M)$ as a function of $m_{\nu_R^M}$ at the LHC with $\sqrt s=$ 8 TeV (top panel) and 13 TeV (bottom panel) for two different values of $\nu_R^M$--$e^M$ mass splitting namely, 50 GeV (left panel) and 100 GeV (right panel). Fig.~\ref{cross_ERNR} also contains SSD signal cross-sections after the acceptance cuts listed in Eqs. (\ref{cut:pT}--\ref{cut:jj-iso}). The corresponding lepton $p_T$ distributions (left panel) and $\not p_T$ distribution (right panel) are presented in Fig.~\ref{dist_ERNR} for the LHC with $\sqrt s=13$ TeV. Large SSD signal cross-sections (see Fig.~\ref{cross_ERNR}) and harder signal leptons $p_T$ and $\not p_T$ distributions (see Fig.~\ref{dist_ERNR}) indicate towards the possibility of detecting SSD signature of {\em EW $\nu_R$} model over the SM background during the ongoing run of LHC with 13 TeV center of mass energy.

In the SM, same sign dilepton arises mainly from the production of $t\bar t W^\pm$ and $ZW^\pm$ productions. $t\bar t W^\pm$ contributes to SSD when $t(\bar t)$ decays leptonically, $\bar t(t)$ decays hadronically and $W^{+(-)}$ decays leptonically. On the other hand, $ZW^\pm$ contributes to SSD when both $Z$ and $W$ decays leptonically and one lepton from $Z$-decay falls out side the coverage of the detector ($p_T<20$ GeV and/or $|\eta|>$2.5) or do not identified as individual entities ($\Delta R_{ll}<0.4$ or $\Delta R_{lj}<0.4$). Contribution to SSD also arises from opposite sign dilepton events due to charge mis-identification. However, the probability of mis-identifying lepton charge is very small. Production of $t\bar t$ pairs also contributes to SSD when  $t\bar t$ pairs decays semileptonically and the $b$-quark from the hadronically decaying top decays into a lepton. However, lepton from the $b$-decay is always accompanied by a lots of hadronic activity around the lepton or a jet within close proximity of the lepton. Therefore, stronger isolation cuts for leptons can be used to eliminate the $t\bar t$ contribution to SSD background. The SM background contribution to SSD was studied by ATLAS collaboration \cite{ATLAS_SSD} in the context of 13 TeV LHC. In order to reduce the SM background contribution to SSD + $\not p_T$, we have used ATLAS suggested cuts on $\not p_T>125$ GeV and $m_{eff}>650$ GeV, where $m_{eff}$ is defined as the scalar sum of the $p_T$ of the signal leptons and jets as well as $\not p_T$, as {\em selection cuts}. With these set of event selection criteria, dominant SM contribution to the SSD arises from $ZW$ and $t\bar tW$ production. We have simulated  $ZW$ and $t\bar tW$ in association with upto 3 and 4 additional jets, respectively, using ALPGEN \cite{Mangano:2002ea} and the resulting SSD background cross-section after the selection cuts is estimated to be 0.6 fb at the LHC with 13 TeV center of mass energy.

\begin{figure}[t]
\includegraphics[scale=0.5]{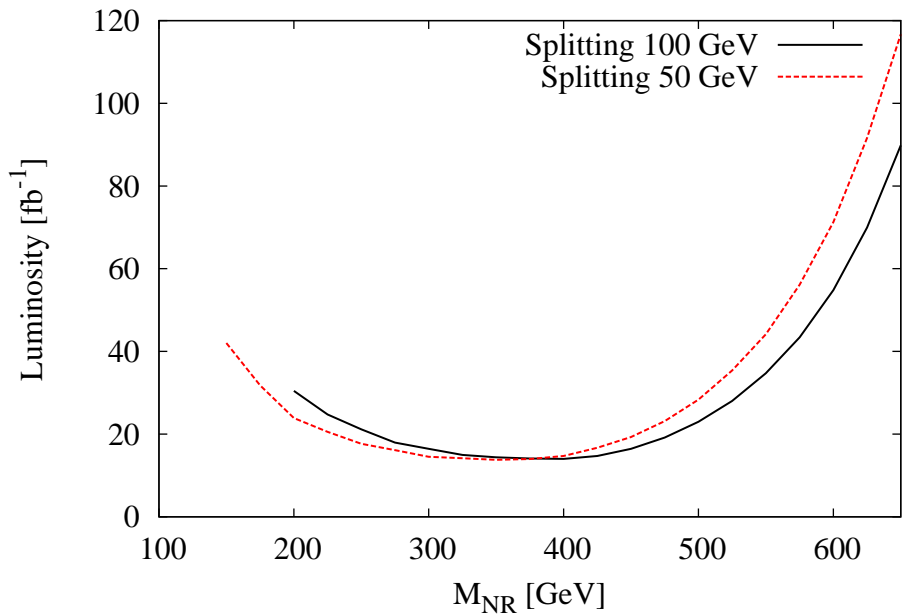} 
\caption{ Required luminosity for 5$\sigma$ discovery at the LHC with $\sqrt s=$13 TeV as a function of right handed mirror neutrino mass for two different values of $\nu_R^M$--$e^M$ mass splitting.}
\label{lumi}
\end{figure}

In order to calculate the discovery reach of the LHC with 13 TeV center of mass energy, we define the signal to be observable for a integrated luminosity ${\cal L}$ if,
\begin{equation}
\frac{N_{S}}{\sqrt{N_B+N_S}} \ge 5,
\end{equation}
where, $N_{S(B)}=\sigma_{S(B)} {\cal L}$, is the number of signal (background) events for an integrated luminosity ${\cal L}$. In Fig.~\ref{lumi}, we have presented required luminosity of the 13 TeV LHC for 5$\sigma$ discovery of $\nu_R^M$ in the framework of {\em EW $\nu_R$} model as a function of $\nu_R^M$ mass. Two lines of Fig.~\ref{lumi} corresponds to two different mass splitting between $\nu_R^M$--$e^M$. Fig.~\ref{lumi} shows that for low mass ($\sim$ 150 GeV) $\nu_R^M$, 5$\sigma$ discovery is possible with about 40 $fb^{-1}$ of integrated luminosity of the LHC running at 13 TeV center of mass energy. To probe intermediate mass range ($\sim$ 200 GeV to 500 GeV), as can be seen from Fig. ~\ref{lumi}, a smaller luminosity ($\sim$ 20 fb$^{-1}$) will suffice. This is a consequence of the hard $\not p_T$ and $m_{eff}$ cuts which apart from reducing the SM background cross-section, also reduces the signal cross-section for low $\nu_R^M$ mass. However, the effect of  hard $\not p_T$ and $m_{eff}$ cuts on the signal cross-section for large $\nu_R^M$ mass is small and with 100 fb$^{-1}$ integrated luminosity we will be able to probe $\nu_R^M$ mass up to 650 (600) GeV for 100 (50) GeV splitting between $\nu_R^M$--$e^M$. 

One important remark is in order here. The analysis presented in this paper concerns mainly with the number of like-sign dileptons regardless of the decay length and the SM background is taken to be those coming from the primary vertex (prompt decays). However, the decay of mirror leptons could be of a displaced-vertex type since the Yukawa couplings that govern the decay rates could be very small as constrained by $\mu \rightarrow e \gamma$. In such a case, the algorithm written for the search will have to be done differently and one cannot simply use the current one. In other words, the analysis presented in this paper can be regarded as the first step in a more complete search for phenomena such as like-sign dileptons coming from the EW $\nu_R$ model.

\section{Summary and Conclusions}

We have presented the LHC phenomenology of the electroweak 
right handed neutrino model. The uniqueness of the model is that the gauge symmetry is the same as the SM, but it has the right handed neutrino, $\nu_R$ as well as the mirror quarks and leptons in the EW scale. The model was invented to explain the tiny neutrino masses with EW scale $\nu_R$ masses. It has one additional Higgs
doublet (called the mirror doublet), two Higgs triplets and four singlet Higgses. The model satisfies the EW precision data as well as all the constraints coming from the 125 GeV Higgs data. One of the interesting features of the model is that the discovered 125 GeV Higgs  has the possibility of coming predominantly from non-SM scalars as explained in the review section above. The model also has interesting prediction for the rare processes such as $\mu \rightarrow e \gamma$ \cite{Hung:2015hra}, $\mu \rightarrow e$ conversion \cite{hung4} which will be explored with a much higher sensitivity in upcoming intensity frontier experiments. 

Because the gauge symmetry is just the $SU(2)_W  \times U(1)_Y$, all the particles get masses from the EW symmetry breaking. As a result, $\nu_R^M$ as well as the mirror quarks and lepton masses can not be larger than a TeV, making ideal for the production of these particle at the LHC. Pair productions of $\nu_R^M$ and the associated productions of $\nu_R^M$ with the charged mirror leptons, $e^M$ are particularly very interesting, because their subsequent decays give rise to the same sign dileptons, and trileptons ($++-$ or $+--$) in the final states. Depending on the mass difference between the $\nu_R^M$ and $e^M$, these final leptons can have high $p_T$ as well as  the events can have large missing energy. Such final states have very small SM background, and relatively few events of this kind will stand out.

In the analysis presented in this paper, we have calculated the pair productions of $\nu_R^M  \nu_R^M$, as well as the associated productions of $\nu_R^M  e_R^M$, (both at $8$ GeV and $13$ GeV LHC) for two different values of the mass splitting between $\nu_R^M$ and $e_R^M$, $50$ GeV and $100$ GeV. Then we looked at how these particles decay, and applied the appropriate LHC cuts  to obtain the signal cross sections for the same sign dilepton final states. The final state cross sections with basic acceptance cuts for the same sign dileptons can be as large as 
$\simeq 100$ fb for the $\nu_R$ mass of $200$ GeV, and decreasing to $\simeq 0.1 fb$ for the $\nu_R$  mass of
$1$ TeV. We have also shown the  $p_T$ distributions of the two same sign dileptons, as well as the missing $p_T$ for the events. These distributions are quite harder, and thus using hard $p_T$ cuts (missing $p_T$ as well as leptons $p_T$) it would be possible to distinguish these events from those coming from the SM. 
We found that for the low mass $\nu_R \sim 150$ GeV, a $5 \sigma$ discovery is possible with $40 fb^{-1}$, whereas for intermediate mass range, $200 - 500$ GeV, a $5 \sigma$ discovery is possible with a lower luminosity $20 fb^{-1}$ as shown in Fig. ~\ref{lumi}. However the reach for higher masses, up to $650$ GeV, a luminosity of $100 fb^{-1}$ would be required. 

Finally we comment that in our analysis, we have assumed that the coupling $g_{sl}$ between the mirror lepton, ordinary lepton and the singlet scalar phi, 
$g_{sl} e_R^M  e_L \phi_S$ is such that the mirror lepton can decay to ordinary lepton and the singlet scalar $\phi_S$ promptly. However, this coupling can be much smaller. The current limit from the rare decay $\mu \rightarrow e \gamma$ is $10^{-3}$ \cite{Hung:2015hra}. If this coupling is much smaller, then the decay will be slow leading to displaced vertices. Such events will have no real SM background, and will be an interesting additional handle to tag such events.


\section*{Acknowledgments}
We are grateful to J. Haley  of ATLAS collaboration and B. Hirosky of the CMS collaboration for several useful discussions regarding the experimental aspects for the detection of our signals. The work of SN was supported in part by a grant from US Department of Energy, Grant Number DE-SC 0010108. PQH and SC acknowledge the support of the Pirrung Foundation. KG is supported by DST (India) under INSPIRE Faculty Award.

\end{document}